\PassOptionsToPackage{table,svgname,hyperref,dvipsnames}{xcolor}
\documentclass[manuscript,nonacm]{acmart}
\pdfoutput=1
\usepackage{pifont}
\usepackage{tikz}
\usepackage{subcaption}
\newcommand{\cmark}{\ding{51}}
\usepackage{bbm}
\usepackage{multirow}
\usepackage{pgfplots}
\usepackage{xspace}
\usepackage{enumitem}
\usepackage{makecell}
\usepackage{adjustbox}

\newcommand{\pbeta}{\ensuremath{\text{RP}^3\!\beta}\xspace}
\newcommand{\easer}{EASE$^R$\xspace}
\newcommand{\userknn}{UserkNN\xspace}
\newcommand{\itemknn}{ItemkNN\xspace}

\usetikzlibrary{positioning,fit,shapes.geometric,backgrounds,calc}
\pgfplotsset{compat=1.18}

\newenvironment{customlegend}[1][]{%
\begingroup
\csname pgfplots@init@cleared@structures\endcsname
\pgfplotsset{#1}%
}{%
\csname pgfplots@createlegend\endcsname
\endgroup
}%
\def\addlegendimage{\csname pgfplots@addlegendimage\endcsname}
\definecolor{forestgreen(web)}{rgb}{0.13, 0.55, 0.13}
\definecolor{lightgray}{rgb}{0.83, 0.83, 0.83}

\AtBeginDocument{%
  \providecommand\BibTeX{{%
    \normalfont B\kern-0.5em{\scshape i\kern-0.25em b}\kern-0.8em\TeX}}}

\definecolor{White}{gray}{0.995}
\usepackage{colortbl}%

\usepackage[table]{xcolor}

\newcolumntype{M}{>{\RaggedRight\arraybackslash}X}

\definecolor{turquoisegreen}{rgb}{0.63, 0.84, 0.71}
\definecolor{uscgold}{rgb}{1.0, 0.8, 0.0}
\definecolor{lightslategray}{rgb}{0.47, 0.53, 0.6}
\definecolor{cinnamon}{rgb}{0.82, 0.41, 0.12}
\definecolor{lincolngreen}{rgb}{0.11, 0.35, 0.02}
\definecolor{mediumcarmine}{rgb}{0.69, 0.25, 0.21}
\definecolor{oceanboatblue}{rgb}{0.0, 0.47, 0.75}
\definecolor{darkorchid}{rgb}{0.6, 0.2, 0.8}
\definecolor{darkkhaki}{rgb}{0.74, 0.72, 0.42}

\copyrightyear{2023}
\acmYear{2023}
\setcopyright{rightsretained}
\acmConference[RecSys '23]{Seventeenth ACM Conference on Recommender Systems}{September 18--22, 2023}{Singapore, Singapore}
\acmBooktitle{Seventeenth ACM Conference on Recommender Systems (RecSys '23), September 18--22, 2023, Singapore, Singapore}\acmDOI{10.1145/3604915.3608759}
\acmISBN{979-8-4007-0241-9/23/09}

\begin{document}

\title[Challenging the Myth of Graph Collaborative Filtering]{Challenging the Myth of Graph Collaborative Filtering: a Reasoned and Reproducibility-driven Analysis}


\author{Vito Walter Anelli}
\authornote{Corresponding authors: Vito Walter Anelli (\url{vitowalter.anelli@poliba.it}),
Daniele Malitesta (\url{daniele.malitesta@poliba.it}),
and Claudio Pomo (\url{claudio.pomo@poliba.it}).}
\email{vitowalter.anelli@poliba.it}
\affiliation{\institution{Politecnico di Bari}
    \country{Italy}
 }

 \author{Daniele Malitesta}
\authornotemark[1]
\email{daniele.malitesta@poliba.it}
\affiliation{\institution{Politecnico di Bari}
\country{Italy}
 }

 \author{Claudio Pomo}
\authornotemark[1]
\email{claudio.pomo@poliba.it}
\affiliation{\institution{Politecnico di Bari, Italy}
\country{Italy}
 }

\author{Alejandro Bellogín}
\email{alejandro.bellogin@uam.es}
 \affiliation{
 \institution{Universidad Autónoma de Madrid}
 \country{Spain}
 }

\author{Eugenio Di Sciascio}
\email{eugenio.disciascio@poliba.it}
\affiliation{\institution{Politecnico di Bari}
\country{Italy}
 }

\author{Tommaso Di Noia}
\email{tommaso.dinoia@poliba.it}
\affiliation{\institution{Politecnico di Bari}
\country{Italy}
 }

\renewcommand{\shortauthors}{Anelli, Malitesta, and Pomo et al.}

\begin{abstract}

The success of graph neural network-based models (GNNs) has significantly advanced recommender systems by effectively modeling users and items as a bipartite, undirected graph. However, many original graph-based works often adopt results from baseline papers without verifying their validity for the specific configuration under analysis. Our work addresses this issue by focusing on the replicability of results. We present a code that successfully replicates results from six popular and recent graph recommendation models (NGCF, DGCF, LightGCN, SGL, UltraGCN, and GFCF) on three common benchmark datasets (Gowalla, Yelp 2018, and Amazon Book). Additionally, we compare these graph models with traditional collaborative filtering models that historically performed well in offline evaluations. Furthermore, we extend our study to two new datasets (Allrecipes and BookCrossing) that lack established setups in existing literature. As the performance on these datasets differs from the previous benchmarks, we analyze the impact of specific dataset characteristics on recommendation accuracy. By investigating the information flow from users' neighborhoods, we aim to identify which models are influenced by intrinsic features in the dataset structure. The code to reproduce our experiments is available at:~\url{https://github.com/sisinflab/Graph-RSs-Reproducibility}.
\end{abstract}



\maketitle

\section{Introduction and related work}

The world of recommender systems (RSs) is experiencing a revolutionary shift, thanks to the emergence of graph neural network-based models~\cite{DBLP:journals/corr/abs-2104-13478, DBLP:journals/tnn/ScarselliGTHM09, DBLP:series/synthesis/2020Hamilton} (GNNs). These groundbreaking models are designed to represent users and items as a bipartite, undirected graph, unlocking a whole new level of high-order relationships that were previously almost unattainable. Not only they do achieve better accuracy than their predecessors, but they are also setting a new standard for modern recommender systems~\cite{DBLP:journals/csur/WuSZXC23, DBLP:conf/wsdm/GaoW0022, DBLP:conf/sigir/0001DWLZ020, DBLP:conf/cikm/MaoZXLWH21}.
In recent years, 
great effort has been devoted in creating 
GNN-based models that address the critical issues of existing 
models, such as the over-smoothing phenomenon~\cite{DBLP:conf/aaai/ChenLLLZS20} and scalability issues~\cite{DBLP:conf/kdd/YingHCEHL18}. These cutting-edge models are taking the world of recommender systems by storm and ushering in a new era of accuracy~\cite{DBLP:conf/cikm/MaoZXLWH21, DBLP:conf/cikm/ShenWZSZLL21, DBLP:conf/cikm/PengSM22, DBLP:journals/corr/abs-2303-08537, DBLP:journals/corr/abs-2302-02113}. 
Over the past ten years, the application of neural techniques rooted in graph representation learning, such as graph convolutional networks~\cite{DBLP:conf/iclr/KipfW17} (GCNs), has introduced a fresh perspective on traditional collaborative filtering (CF) approaches. Rather than relying solely on user-item interactions for optimization~\cite{DBLP:journals/computer/KorenBV09, DBLP:conf/uai/RendleFGS09, DBLP:conf/www/HeLZNHC17}, GCN-based methods enable the extraction of both short- and long-distance user preferences toward items~\cite{DBLP:conf/sigir/Wang0WFC19}. By incorporating multi-hop relationships into the embeddings of users and items, these learned profiles yield more precise recommendations, as evidenced in the literature~\cite{DBLP:conf/sigir/0001DWLZ020, DBLP:conf/cikm/MaoZXLWH21}. Nevertheless, more researchers obtained different accuracy outcomes in independent experiments and began questioning the graph collaborative filtering (graph CF) prominence~\cite{DBLP:conf/sigir/ZhuDSMLCXZ22}.

The original GCN layer employs message-passing techniques to refine the node representations of users and items through the iterative aggregation of their respective multi-hop neighbor nodes. While early attempts focused on simple aggregation methods~\cite{DBLP:journals/corr/BergKW17, DBLP:conf/kdd/YingHCEHL18}, recent solutions have advanced the field by exploring the inter-dependencies between nodes and their neighbors~\cite{DBLP:conf/sigir/Wang0WFC19}, designing simplified versions of the graph convolutional layer~\cite{DBLP:conf/sigir/0001DWLZ020, DBLP:conf/aaai/ChenWHZW20} and learning multiple nodes' views~\cite{DBLP:conf/sigir/WuWF0CLX21, DBLP:conf/sigir/YuY00CN22} augmented via self-supervised and contrastive learning to improve model accuracy. Moreover, current trends aim to simplify message-passing formulations~\cite{DBLP:conf/cikm/MaoZXLWH21, DBLP:conf/cikm/ShenWZSZLL21, DBLP:conf/cikm/PengSM22}, explore other spaces for graph-based recommendation tasks~\cite{DBLP:conf/cikm/ShenWZSZLL21, DBLP:conf/www/SunCZPV21, DBLP:conf/sigir/ZhangL0WSLSZDZ22}, and use hypergraphs to capture complex user-item dependencies~\cite{DBLP:conf/sigir/XiaHXZYH22, DBLP:conf/cikm/WeiLBL22}.
To filter out noisy neighbors and uncover hidden preference patterns, a complementary research field emerged that focuses on learning importance weights through attention mechanisms, such as those employed in the graph attention network~\cite{DBLP:conf/iclr/VelickovicCCRLB18} (GAT). While some models aim to recognize meaningful user-item interactions at a higher level~\cite{DBLP:conf/kdd/Wang00LC19, DBLP:journals/ipm/TaoWWHHC20}, others disentangle relations on a finer-grained scale~\cite{DBLP:conf/sigir/WangJZ0XC20, DBLP:conf/sigir/ZhangL0WSLSZDZ22}. The recent advancements in GCN-based techniques have opened up new avenues for more accurate and effective recommendation systems.

Reproducibility is the cutting-edge research task in which researchers replicate experimental results using the same data and methods~\cite{DBLP:journals/tois/DacremaBCJ21, DBLP:journals/umuai/BelloginS21, DBLP:conf/recsys/SunY00Q0G20, DBLP:conf/recsys/DacremaCJ19}. In the case of graph CF, several factors contribute to the lack of reproducibility. Firstly, many graph CF studies copy previous results found in the literature for the same datasets, which makes it challenging to compare and reproduce results across different studies. Secondly, such studies 
do not provide the implementation of the adopted baselines, which makes it difficult to assess the effectiveness of different models.
Furthermore, graph CF studies frequently do not provide complete information since they do not always share the experimental setups, such as hyper-parameter settings and training procedures. 
This lack of transparency makes it challenging to reproduce results and verify the validity of the findings.
The lack of reproducibility in graph CF is a significant issue because it undermines the research's credibility and hinders the field's progress. To address this problem, researchers should strive to provide more detailed descriptions of their experimental setups and make their code and datasets publicly available. Additionally, the research community should work together to establish standard evaluation metrics and experimental protocols to promote reproducibility and facilitate comparison across different studies.

To this aim, this work reports on a notable reproducibility effort to re-implement and replicate the results of six state-of-the-art (both well-established and recent) papers on graph collaborative filtering, namely, NGCF~\cite{DBLP:conf/sigir/Wang0WFC19}, DGCF~\cite{DBLP:conf/sigir/WangJZ0XC20}, LightGCN~\cite{DBLP:conf/sigir/0001DWLZ020},
SGL~\cite{DBLP:conf/sigir/WuWF0CLX21}, UltraGCN~\cite{DBLP:conf/cikm/MaoZXLWH21}, and GFCF~\cite{DBLP:conf/cikm/ShenWZSZLL21}.
In particular, we provide an in-depth experimental analysis of the papers, conducting the experiments from scratch on the three 
datasets adopted in the original papers: Gowalla~\cite{DBLP:conf/www/LiangCMB16}, Yelp 2018~\cite{DBLP:conf/sigir/0001DWLZ020}, and Amazon Book~\cite{DBLP:conf/www/HeM16}.
Notably, the investigation extends the previous works by incorporating state-of-the-art classical collaborative filtering baselines such as \userknn~\cite{DBLP:conf/cscw/ResnickISBR94}, \itemknn~\cite{DBLP:conf/www/SarwarKKR01}, \pbeta~\cite{DBLP:journals/tiis/PaudelCNB17}, and \easer~\cite{DBLP:conf/www/Steck19} to correctly position the graph CF methods in the recommender systems state-of-the-art.

The study's findings reveal that \pbeta ranks as the second-best method with the Yelp 2018 dataset, indicating that the original papers would have needed a more comprehensive evaluation. To this end, the evaluation benchmark incorporates two additional datasets, Allrecipes~\cite{DBLP:journals/tmm/GaoFHHGFMC20} and BookCrossing~\cite{DBLP:conf/www/ZieglerMKL05}, which are common in the recommendation literature but uncommon in the graph CF-specific literature. However, surprisingly, the rankings significantly differ on the Allrecipes dataset, and the mathematical formulation of the graph CF methods is not sufficient to account for these outcomes. This observation leads to further investigation to comprehend the experimental results.
Examining the dataset \textit{topological} characteristics shows that the overall number of users and items and the average user and item degree vary from dataset to dataset. This observation may indicate the amount of information transmitted from node to node in the computational graph. According to the mathematical background, the analysis of the results is then threefold, focusing on the impact of (i) the \textit{coldness/warmness} of a user, (ii) the \textit{popularity} of the enjoyed items, and (iii) the \textit{size} of the user \textit{neighborhood} and the coldness/warmness of the neighbors. The users are partitioned in quartiles accordingly, and the experiments are re-evaluated to obtain more fine-grained results that motivate the outcomes for all the considered datasets.

Overall, the study aims to comprehensively answer several research questions, including:
\begin{itemize}[leftmargin=1.2cm]
    \item[\textbf{RQ1.}] Is the state-of-the-art (i.e., the six most important papers) of graph collaborative filtering (graph CF) replicable?
    \item[\textbf{RQ2.}] How does the state-of-art of graph CF position with respect to classic CF state-of-the-art?
    \item[\textbf{RQ3.}] How does the state-of-art of graph CF perform on datasets from different domains and with different topological aspects, not commonly adopted for graph CF recommendation?
    \item[\textbf{RQ4.}] What information (or lack of it) impacts the performance of the graph CF methods across the various datasets?
\end{itemize}

The following introduces the background and the experiments to answer the outlined research questions. First, in~Section \ref{sec:background}, we present the background technologies and the reproducibility details to conduct our study. Then, in~Section \ref{sec:prior-results}, we report the reproducibility results, whose insights are complemented by adding novel classic CF baselines (i.e.,~Section \ref{sec:other-baselines}). Furthermore, an investigation upon graph topology sheds light on the discrepancies of the graph CF approaches on two introduced datasets (i.e.,~Section \ref{sec:other-datasets}). By reinterpreting the concept of users' node degree as \textit{information flow} from the multi-hop neighborhoods to the user, we unveil the behavior of the graph and classic CF. Finally, Section \ref{sec:conclusion} wraps up the main take-home messages and paves the way to novel directions for future work. Codes and datasets to reproduce our paper are available here:~\url{https://github.com/sisinflab/Graph-RSs-Reproducibility}.

\section{Background and reproducibility analysis} 
\label{sec:background}

The current section is aimed to provide the background about selected state-of-the-art methodologies in 
graph CF
and their reproducibility details as presented in the original papers. First, the main aspects about graph-based models are introduced to conduct a chronological analysis of the strategies behind each algorithm (Section \ref{sec:graph-collab-filtering}). Then, we assess the experimental settings as reported in the original works by focusing on the chosen baselines (Section \ref{sec:baselines}), the datasets involved (Section \ref{sec:datasets}), and the training-testing protocol adopted in each case (Section \ref{sec:comparison}).

\subsection{Graph collaborative filtering} \label{sec:graph-collab-filtering}

In graph CF, users, items, and their interconnections are viewed as a bipartite and undirected graph. Let $\mathcal{U}$ and $\mathcal{I}$ be the sets of users and items in the recommendation system, respectively. Then, let $\mathbf{R} \in \mathbb{R}^{|\mathcal{U}| \times |\mathcal{I}|}$ be the user-item interaction matrix where, in an implicit feedback scenario, $R_{ui} = 1$ if user $u \in \mathcal{U}$ interacted with item $i \in \mathcal{I}$, 0 otherwise. We build the adjacency matrix $\mathbf{A} \in \mathbb{R}^{(|\mathcal{U}| + |\mathcal{I}|) \times (|\mathcal{U}| + |\mathcal{I}|)}$ indicating the bi-directional connections linking users and items in $\mathbf{R}$:
\begin{equation}
    \mathbf{A} = \begin{bmatrix}
    0 & \mathbf{R} \\
    \mathbf{R}^\top & 0 
    \end{bmatrix}.
\end{equation}
We use the set of users and items, along with the adjacency matrix, to formally define the user-item bipartite and undirected graph $\mathcal{G} = \{\mathcal{U} \cup \mathcal{I}, \mathbf{A}\}$. 
By associating users' and items' nodes to embeddings, the vast majority of approaches iteratively update their representations at different hop distances through the message-passing schema~\cite{DBLP:conf/icml/GilmerSRVD17, DBLP:journals/corr/abs-2104-13478}. 

For this work, we select and reproduce the results for six widely-recognized state-of-the-art approaches in graph CF, namely, NGCF \cite{DBLP:conf/sigir/Wang0WFC19}, DGCF~\cite{DBLP:conf/sigir/WangJZ0XC20}, LightGCN~\cite{DBLP:conf/sigir/0001DWLZ020}, SGL~\cite{DBLP:conf/sigir/WuWF0CLX21}, UltraGCN~\cite{DBLP:conf/cikm/MaoZXLWH21}, and GFCF~\cite{DBLP:conf/cikm/ShenWZSZLL21} (refer to~Section \ref{sec:prior-results}). 
This 
selection is motivated by two aspects: (i) such models are adopted as baselines in recent works from top-tier venues (see the second column in Table \ref{tab:models}); (ii) their strategies cover a wide spectrum of techniques in graph CF. To provide a chronological overview of such techniques, in the following, we report their main aspects: 
\begin{itemize}[leftmargin=*]
\item \textbf{NGCF.} Neural graph collaborative filtering~\cite{DBLP:conf/sigir/Wang0WFC19} (NGCF) is among the pioneer approaches in graph CF. Its message-passing schema works by aggregating the neighborhood information and the inter-dependencies among the \textit{ego} and the \textit{neighborhood} nodes 
(note that a normalized Laplacian adjacency matrix is used during the message-passing).

\item \textbf{DGCF.} Disentangled graph collaborative filtering~\cite{DBLP:conf/sigir/WangJZ0XC20} (DGCF) assumes that user-item interactions can be disentangled into independent intents, where each stands for a specific aspect describing the user's preference towards the item. The model learns a set of weighted adjacency matrices 
refining the user-item importance related to a specific intent. 

\item \textbf{LightGCN.} Light graph convolutional network~\cite{DBLP:conf/sigir/0001DWLZ020} (LightGCN) 
suggests that a more light-weight formulation of 
the graph convolutional layer proposed by~\citet{DBLP:conf/iclr/KipfW17} can lead to superior accuracy performance in the recommendation scenario. Specifically, the architecture removes feature transformations and non-linearities.

\item \textbf{SGL.} Self-supervised graph learning~\cite{DBLP:conf/sigir/WuWF0CLX21} (SGL) is among the first attempts to bring the lesson-learned from self-supervised~\cite{DBLP:conf/cikm/HuangXW0Y22} and contrastive~\cite{DBLP:conf/nips/KhoslaTWSTIMLK20} learning to graph CF. Built upon a LightGCN-based convolutional layer, the model learns different views of nodes by performing node/edge dropout and random walk operations on the graph topology. A self-supervised contrastive loss component is added to encourage the consistency among different views of the same node and the divergence among different nodes.

\item \textbf{UltraGCN.} Ultra simplification of graph convolutional network \cite{DBLP:conf/cikm/MaoZXLWH21} (UltraGCN) addresses some crucial issues in graph CF. Specifically, the authors propose a novel message-passing schema that mathematically approximates the infinite-layer propagation through a single (simplified) node update iteration.
The adjacency matrix is normalized through a modified Laplacian formulation that accounts for the asymmetric weighting of connected nodes in user-user and item-item connections. Moreover, two loss components are introduced to tackle the over-smoothing effect and learn from the usually-unexplored type of node relationships such as item-item.

\item \textbf{GFCF.} Graph filter-based collaborative filtering~\cite{DBLP:conf/cikm/ShenWZSZLL21} (GFCF) questions the role of graph convolutional network into recommendation by leveraging graph signal processing theory. By showing that several existing approaches in CF may fall into one unified framework based upon graph convolution, the authors eventually propose a closed-form algorithm that proves to be a strong baseline against other trainable and computationally-expensive (graph-based) approaches in CF. Thus, the method represents the only exception to the message-passing models presented above.
\end{itemize}

\subsection{Analysis on reported baselines} \label{sec:baselines}
Table \ref{tab:models} reports on the baselines each graph-based approach was tested against in the original paper. By categorizing them into \textit{classic} and \textit{graph} CF we first observe that, with the only exception of UltraGCN, all graph-based recommendation systems are generally compared only against 1-2 classical CF solutions (MF~\cite{DBLP:conf/uai/RendleFGS09, DBLP:conf/www/HeLZNHC17}- and/or VAE~\cite{DBLP:conf/www/LiangKHJ18, DBLP:conf/nips/MaZ0Y019}-based approaches in most cases). However, the recent literature~\cite{DBLP:conf/recsys/DacremaCJ19, DBLP:journals/tois/DacremaBCJ21, DBLP:conf/recsys/AnelliBNP21, DBLP:conf/um/AnelliBNJP22, DBLP:conf/sigir/ZhuDSMLCXZ22} has raised several concerns about usually-untested strong CF baselines, such as nearest-neighborhood approaches (e.g., \userknn~\cite{DBLP:conf/cscw/ResnickISBR94} and \itemknn~\cite{DBLP:conf/www/SarwarKKR01}), random-walk techniques (e.g., \pbeta~\cite{DBLP:journals/tiis/PaudelCNB17}), and other autoencoder-based solutions (e.g., \easer~\cite{DBLP:conf/www/Steck19}). 
Differently from the classical CF baselines, we notice that most of the works compare their proposed approaches against a wide (and shared) range of graph CF solutions. This is easily explainable given the conceptual and logical similarities among the graph CF baselines and the proposed approaches. Moreover, besides a limited subset of graph CF baselines (i.e., HOP-Rec~\cite{DBLP:conf/recsys/YangCWT18} and GRMF~\cite{DBLP:conf/nips/RaoYRD15}), the vast majority of tested graph algorithms~\cite{DBLP:journals/corr/BergKW17, DBLP:conf/kdd/YingHCEHL18, DBLP:conf/icml/Ma0KW019, DBLP:conf/sigir/SunZGGTHMC20, DBLP:conf/aaai/ChenWHZW20} are based upon the graph convolutional network architecture. Interestingly, we observe that only a subgroup of our selected six graph CF approaches (up to a maximum of three approaches if we consider UltraGCN) is generally compared against the proposed approach. While we could justify this point with chronological motivations (e.g., DGCF could have not been tested on SGL, UltraGCN, and GFCF), we deem this to be an important lack in the existing literature.

\begin{table*}[!t]
\centering
\caption{Analysis of baselines used in each of the selected graph-based models, categorized into \textit{classic} and \textit{graph} CF. A colored tick `\textcolor{orange}{\cmark}' denotes when one of the baselines is also among the selected set of graph-based approaches for our study.}
\label{tab:models}
\footnotesize
\begin{adjustbox}{width=\textwidth, center}
\setlength{\tabcolsep}{4pt}
\begin{tabular}{llcccccc}
\toprule
\multirow{4}{*}{\textbf{Families}} & \multirow{4}{*}{\textbf{Baselines}} & 
\multicolumn{6}{c}{\textbf{Models}} \\ \cmidrule{3-8} & & NGCF~\cite{DBLP:conf/sigir/Wang0WFC19} & DGCF~\cite{DBLP:conf/sigir/WangJZ0XC20} & LightGCN~\cite{DBLP:conf/sigir/0001DWLZ020} & SGL~\cite{DBLP:conf/sigir/WuWF0CLX21} & UltraGCN~\cite{DBLP:conf/cikm/MaoZXLWH21} & GFCF~\cite{DBLP:conf/cikm/ShenWZSZLL21} \\ \cmidrule{3-8}
& & \multicolumn{6}{c}{\textbf{Used as graph CF baseline in (2021 --- present)}} \\ \cmidrule{3-8} & & \cite{DBLP:conf/kdd/HuangDDYFW021, DBLP:conf/wsdm/CaoLGLL021, DBLP:conf/www/SunCZPV21, DBLP:conf/wsdm/WeiHXXZY22, DBLP:conf/www/000100LK22, DBLP:conf/wsdm/Chen0S0ZZ22} & \cite{DBLP:conf/cikm/MaoZWDDXH21, DBLP:conf/sigir/ZhangL0WSLSZDZ22, DBLP:conf/sigir/FanL0ZT022, 9736612, DBLP:conf/wsdm/WangZS22, DBLP:conf/www/LinTHZ22} & \cite{DBLP:conf/sigir/WuWF0CLX21, DBLP:conf/www/YuYLWH021, DBLP:conf/www/LiuCZGN21, DBLP:conf/sigir/YuY00CN22, DBLP:conf/sigir/XiaHXZYH22, DBLP:conf/kdd/RaoCLSYH22} & \cite{DBLP:conf/cikm/MaoZWDDXH21, DBLP:conf/sigir/YangHXL22, DBLP:conf/sigir/Gao0HCZFZ22, DBLP:conf/sigir/0002WLCZDWSLW22, DBLP:conf/wsdm/WeiHXXZY22, DBLP:conf/sigir/XiaHXZYH22} & \cite{DBLP:conf/mm/DuWF0022, DBLP:conf/cikm/GongSWLL22, DBLP:conf/sigir/ZhuDSMLCXZ22, DBLP:conf/cikm/OuyangWP22, DBLP:conf/sigir/0001OM22, 10.1145/3570501} & \cite{DBLP:conf/recsys/AnelliDNSFMP22, DBLP:conf/sigir/ZhuDSMLCXZ22, DBLP:conf/sigir/PengSM22, DBLP:conf/www/XiaLGLLG22, DBLP:journals/corr/abs-2302-02113, DBLP:conf/ecir/AnelliDNMPP23} \\ \cmidrule{1-8}
\multirow{15}{*}{\textit{Classic CF}} & MF-BPR~\cite{DBLP:conf/uai/RendleFGS09} & \cmark & \cmark &  & & \cmark \\ \cmidrule{2-8}
& NeuMF~\cite{DBLP:conf/www/HeLZNHC17} & \cmark \\ \cmidrule{2-8}
& CMN~\cite{DBLP:conf/sigir/EbesuSF18} & \cmark \\ \cmidrule{2-8}
& MacridVAE~\cite{DBLP:conf/nips/MaZ0Y019} &  & \cmark \\ \cmidrule{2-8}
& Mult-VAE~\cite{DBLP:conf/www/LiangKHJ18} &  &  & \cmark & \cmark &  & \cmark \\ \cmidrule{2-8}
& DNN+SSL~\cite{DBLP:journals/corr/abs-2007-12865} &  &  &  & \cmark \\ \cmidrule{2-8}
& ENMF~\cite{DBLP:journals/tois/ChenZZLM20} &  &  &  &  &  \cmark \\ \cmidrule{2-8}
& CML~\cite{DBLP:conf/www/HsiehYCLBE17} &  &  &  &  &  \cmark \\ \cmidrule{2-8}
& DeepWalk~\cite{DBLP:conf/kdd/PerozziAS14} &  &  &  &  &  \cmark \\ \cmidrule{2-8}
& LINE~\cite{DBLP:conf/www/TangQWZYM15} &  &  &  &  &  \cmark \\ \cmidrule{2-8}
& Node2Vec~\cite{DBLP:conf/kdd/GroverL16} &  &  &  &  &  \cmark \\ \cmidrule{2-8}
& NBPO~\cite{DBLP:conf/sigir/YuQ20} &  &  &  &  &  \cmark \\ \midrule
\multirow{17}{*}{\textit{Graph CF}} & HOP-Rec~\cite{DBLP:conf/recsys/YangCWT18} & \cmark \\ \cmidrule{2-8}
& GC-MC~\cite{DBLP:journals/corr/BergKW17} & \cmark & \cmark \\ \cmidrule{2-8}
& PinSage~\cite{DBLP:conf/kdd/YingHCEHL18} & \cmark \\ \cmidrule{2-8}
& NGCF~\cite{DBLP:conf/sigir/Wang0WFC19} &  & \textcolor{orange}{\cmark} & \textcolor{orange}{\cmark} & \textcolor{orange}{\cmark} & \textcolor{orange}{\cmark} & \textcolor{orange}{\cmark} \\ \cmidrule{2-8}
& DisenGCN~\cite{DBLP:conf/icml/Ma0KW019} &  & \cmark \\ \cmidrule{2-8}
& GRMF~\cite{DBLP:conf/nips/RaoYRD15} &  &  &  \cmark &  &  &  \cmark \\ \cmidrule{2-8}
& GRMF-Norm~\cite{DBLP:conf/sigir/0001DWLZ020} &  &  &  \cmark &  &  &  \cmark \\ \cmidrule{2-8}
& NIA-GCN~\cite{DBLP:conf/sigir/SunZGGTHMC20} &  &  &  &  &  \cmark \\ \cmidrule{2-8}
& LightGCN~\cite{DBLP:conf/sigir/0001DWLZ020} &  &  &  & \textcolor{orange}{\cmark} & \textcolor{orange}{\cmark} & \textcolor{orange}{\cmark} \\ \cmidrule{2-8}
& DGCF &  &  &  &  & \textcolor{orange}{\cmark} \\ \cmidrule{2-8}
& LR-GCCF~\cite{DBLP:conf/aaai/ChenWHZW20} &  &  &  &  &  \cmark \\ \cmidrule{2-8}
& SCF~\cite{DBLP:conf/recsys/ZhengLJZY18} &  &  &  &  &  \cmark \\ \cmidrule{2-8}
& BGCF~\cite{DBLP:conf/kdd/SunGZZRHGTYHC20} &  &  &  &  &  \cmark \\ \cmidrule{2-8}
& LCFN~\cite{DBLP:conf/icml/YuQ20} &  &  &  &  &  \cmark \\
\bottomrule
\end{tabular}
\end{adjustbox}
\end{table*}

Under the above considerations, and differently from the previous works, we compare the accuracy performance of the selected six graph CF approaches against strong CF techniques (\userknn, \itemknn, \pbeta and \easer), while providing a complete evaluation setting which involves all the selected graph methods, where they are put against one another (refer to~Section \ref{sec:other-baselines}). To our knowledge, this work is one of the first attempts~\cite{DBLP:conf/sigir/ZhuDSMLCXZ22} to fill this gap. 

\subsection{Analysis on reported datasets}\label{sec:datasets}
Table \ref{tab:datasets} displays the datasets adopted to train and test the reviewed graph-based recommender systems, as reported in the original papers. Notably, we recognize a total of seven recommendation datasets spanning different domains such as social networks (i.e., Gowalla), points-of-interest (i.e., Yelp 2018), e-commerce (i.e., the Amazon product categories and Alibaba-iFashion), and movies (i.e., Movielens 1M). It is worth pointing out that when we set the `\cmark' for the same dataset on different models, we are stating that the authors from the original works used the exact same dataset setting, that is, the original user-item interaction data and splitting/filtering strategies. 
A deeper analysis shows that there exists a subset of three datasets (i.e., Gowalla~\cite{DBLP:conf/www/LiangCMB16}, Yelp 2018~\cite{DBLP:conf/sigir/0001DWLZ020}, and Amazon Book~\cite{DBLP:conf/www/HeM16}) which is utilized in the majority of graph CF works. For the sake of reproducibility, we replicate the original results calculated on such datasets for the six graph CF approaches (although the SGL paper does not provide results on Gowalla). Given the limited set of shared datasets among all the approaches, we include novel, \textit{never-investigated} datasets to assess if their recommendation accuracy remains consistent on other domains and/or topologies.

\begin{table*}[!t]
\centering
\caption{Analysis of the datasets adopted in each graph-based approach.}
\label{tab:datasets}
\scriptsize
\begin{tabular}{lccccccc}
\toprule
\textbf{Models} & \textbf{Gowalla} & \textbf{Yelp 2018} & \textbf{Amazon Book} & \textbf{Alibaba-iFashion} & \textbf{Movielens 1M} & \textbf{Amazon Electronics} & \textbf{Amazon CDs} \\ \cmidrule{1-8} 
NGCF & \cmark & \cmark & \cmark \\ \cmidrule{1-8}
DGCF & \cmark & \cmark & \cmark \\ \cmidrule{1-8}
LightGCN & \cmark & \cmark & \cmark \\ \cmidrule{1-8}
SGL &  & \cmark & \cmark & \cmark \\ \cmidrule{1-8}
UltraGCN & \cmark & \cmark & \cmark &  & \cmark & \cmark & \cmark \\ \cmidrule{1-8}
GFCF & \cmark & \cmark & \cmark \\
\bottomrule
\end{tabular}
\end{table*}

\subsection{Analysis on experimental comparison} \label{sec:comparison}
As a final analyzed dimension, we discuss the protocol for the experimental comparison between the baselines and the proposed approach in each selected work. Being the pioneer model in the domain, the authors from NGCF train all proposed baselines from scratch. In the DGCF paper, the authors directly report the results of some baselines which are shared with NGCF and train the other baselines from scratch. In a similar manner, the authors by LightGCN, SGL, and UltraGCN copy the result values from the original papers, while the remaining models are trained from scratch. Finally, the authors from GFCF reproduce the results from LightGCN as the baselines are exactly the same. 


Regarding the copy-paste of baseline results, authors often justify this approach by stating that they used the same experimental settings (such as dataset splitting/filtering) as their (graph) CF baselines. Additionally, it is worth noting that some authors are shared among the studies being investigated.

To remove all doubts, and differently from the mentioned works, we re-implement all algorithms by carefully following their original codes, and train/evaluate them through Elliot~\cite{DBLP:conf/sigir/AnelliBFMMPDN21,DBLP:conf/um/MalitestaPANF23}. Our goal is to provide a fair and repeatable experimental environment for the selected graph CF approaches, by using the hyper-parameter settings as indicated in each paper and/or shared online code to assess to what extent we can reproduce the original results.
The reader may refer to~Section \ref{replic_set} for a whole description of our settings. 





\section{Replication of prior results (RQ1)}
\label{sec:prior-results}
This section focuses on how the replication of the experiments from  the six state-of-the-art papers on graph CF stated before has been set up. It starts by defining the evaluation protocol applied to compare these methods in their respective works (Section~\ref{replic_set}). After that, we present our replication results (Section~\ref{replic_res}).

\subsection{Settings}
\label{replic_set}
The experimental setup adopted in the first part of this study is designed primarily to replicate the results of the models included in this analysis~\cite{DBLP:conf/sigir/Wang0WFC19,DBLP:conf/sigir/0001DWLZ020,DBLP:conf/icml/Ma0KW019,DBLP:conf/cikm/HuangXW0Y22,DBLP:conf/cikm/MaoZXLWH21,DBLP:conf/cikm/ShenWZSZLL21}. As mentioned earlier, we use the three most common datasets in this scenario to show the results of our replicability study. Specifically, we use Gowalla, Yelp 2018, and Amazon Book as provided in the public
repositories of NGCF\footnote{\url{https://github.com/xiangwang1223/neural_graph_collaborative_filtering}.} and LightGCN\footnote{\url{https://github.com/kuandeng/LightGCN}.}. All the proposed models (except SGL) use the same datasets with the same filtering/splitting. The authors state that they adopt a random split based on the 80/20 hold-out (i.e., for each user, 80\% of the interactions is used to create the training set, while the remaining 20\% constitutes the test set). Thus, each user-item interaction is treated as positive; all others are considered unfavorable. In addition, the authors leave 10\% of the training as a validation set for tuning the hyper-parameters. However, this portion of the dataset is not indicated in the papers' extra material.

The adopted evaluation protocol, all-unrated-item~\cite{DBLP:conf/recsys/Steck13}, is shared across all the analyzed papers: for each user, we retain all candidate items with whom she does not interact with in the training set. The quality of recommendations is assessed by the Recall and the nDCG on the top-20 recommendation lists for each user.
Each work performs its own tuning of the hyper-parameters (the Recall@20 is used as validation metric), by reporting on the search hyper-parameter spaces. Moreover, the best configurations on each dataset are usually provided in the respective papers and/or repositories.
Thus, we set the hyper-parameters on each model-dataset as the best ones declared by the authors.
The careful reader would notice that the results reported in~Table \ref{tab:reproducibility} for NGCF (see the `\textbf{Original}' column) differ from those shown in the in-proceedings version~\cite{DBLP:conf/sigir/Wang0WFC19}.
The reason is that the authors modified and recalculated the results obtained for the model and baselines due to errors in the pre-processing of the Yelp 2018 dataset and in the calculation of the nDCG. Thus, for the sake of fair reproducibility, and only in this case, we consider the results reported in the arXiv (most updated) version of the paper~\cite{DBLP:journals/corr/abs-1905-08108}.

\subsection{Results}
\label{replic_res}

Table~\ref{tab:reproducibility} compares the results reported by the six papers focused on our study with those obtained in our implementation (using the tuned parameters specified in each work, as explained before).
The new experiments closely approximate the original ones, with the most significant performance shift being in the $10^{-3}$ order. There are no noticeable distinctions in metrics, dataset, or algorithm used.

More specifically, in an algorithm basis, we observe
the performance of GFCF is the best replicated one.
This might be due to 
this method being the only one with a closed-form algorithm, hence, no perturbations from random initializations are expected.
The rest of the approaches evidence a similar (high) level of replication, although the shift for NGCF and DGCF rarely achieves the $10^{-4}$ order for the two metrics in all the datasets. In any case, considering
the random initializations and stochastic learning processes \cite{jordan2023calibrated},
our replication of these approaches could be considered a success.

No significant differences were found among the three datasets. SGL was not originally reported for Gowalla, so it was omitted from the table as we compared reported results with our implementations using the same hyper-parameters.

In summary, these results confirm that, as discussed before,
even though authors of these papers re-used the performance values from other papers just by copy-pasting them, this did not hurt the reproducibility of these approaches.
As previously stated, our assumption for this behavior (which is not a safe practice in general~\cite{DBLP:journals/umuai/BelloginS21}) is that the experiments of the original papers were all comparable because some authors are shared across contributions, which should guarantee that the settings and implementations of the algorithms are the same.

\begin{table*}[!t]
\centering
\caption{Results of our replicability study on Gowalla, Yelp 2018, and Amazon Book for the selected state-of-the-art graph-based recommender systems. We calculate the performance shift between our conducted experiments and the original ones (as reported in their papers). Note that models have been sorted out according to the chronological order.}
\label{tab:reproducibility}
\footnotesize
\begin{tabular}{llcccccc}
\toprule
\textbf{Datasets} & \textbf{Models} & \multicolumn{2}{c}{\textbf{Ours}} & \multicolumn{2}{c}{\textbf{Original}} & \multicolumn{2}{c}{\textbf{Performance Shift}} \\ \cmidrule{3-8} 
\multicolumn{2}{c}{} & \multicolumn{1}{c}{Recall} & \multicolumn{1}{c}{nDCG} & \multicolumn{1}{c}{Recall} & \multicolumn{1}{c}{nDCG} & \multicolumn{1}{c}{Recall} & \multicolumn{1}{c}{nDCG} \\ \cmidrule{1-8} 
\multirow{6}{*}{\textbf{Gowalla}} & NGCF & 0.1556 & 0.1320 & 0.1569 & 0.1327 & \multicolumn{1}{r}{$-1.3\cdot10^{-03}$} & \multicolumn{1}{r}{$-7\cdot10^{-04}$}\\
& DGCF & 0.1736 & 0.1477 & 0.1794 & 0.1521 & \multicolumn{1}{r}{$-5.8\cdot10^{-03}$} & \multicolumn{1}{r}{$-4.4\cdot10^{-03}$}\\
& LightGCN & 0.1826 & 0.1545 & 0.1830 & 0.1554 & \multicolumn{1}{r}{$-4\cdot10^{-04}$} & \multicolumn{1}{r}{$-9\cdot10^{-04}$}\\
& SGL* & --- & --- & --- & --- & \multicolumn{1}{r}{---} & \multicolumn{1}{r}{---} \\
& UltraGCN & 0.1863 & 0.1580 & 0.1862 & 0.1580 & \multicolumn{1}{r}{$+1\cdot10^{-04}$} & \multicolumn{1}{r}{0} \\
& GFCF & 0.1849 & 0.1518 & 0.1849 & 0.1518 & \multicolumn{1}{r}{0} & \multicolumn{1}{r}{0} \\ \cmidrule{1-8}
\multirow{6}{*}{\textbf{Yelp 2018}} & NGCF & 0.0556 & 0.0452 & 0.0579 & 0.0477 & \multicolumn{1}{r}{$-2.3\cdot10^{-03}$} & \multicolumn{1}{r}{$-2.5\cdot10^{-03}$}\\
& DGCF & 0.0621 & 0.0505 & 0.0640 & 0.0522 & \multicolumn{1}{r}{$-1.9\cdot10^{-03}$} & \multicolumn{1}{r}{$-1.7\cdot10^{-03}$} \\
& LightGCN & 0.0629 & 0.0516 & 0.0649 & 0.0530 & \multicolumn{1}{r}{$-2\cdot10^{-03}$} & \multicolumn{1}{r}{$-1.4\cdot10^{-03}$} \\
& SGL & 0.0669 & 0.0552 & 0.0675 & 0.0555 & \multicolumn{1}{r}{$-6\cdot10^{-04}$} & \multicolumn{1}{r}{$-3\cdot10^{-04}$} \\
& UltraGCN & 0.0672 & 0.0553 & 0.0683 & 0.0561 & \multicolumn{1}{r}{$-1.1\cdot10^{-03}$} & \multicolumn{1}{r}{$-8\cdot10^{-04}$} \\
& GFCF & 0.0697 & 0.0571 & 0.0697 & 0.0571 & \multicolumn{1}{r}{0} & \multicolumn{1}{r}{0} \\ \cmidrule{1-8}
\multirow{6}{*}{\textbf{Amazon Book}} & NGCF & 0.0319 & 0.0246 & 0.0337 & 0.0261 & \multicolumn{1}{r}{$-1.8\cdot10^{-03}$} & \multicolumn{1}{r}{$-1.5\cdot10^{-03}$}\\
& DGCF & 0.0384 & 0.0295 & 0.0399 & 0.0308 & \multicolumn{1}{r}{$-1.5\cdot10^{-03}$} & \multicolumn{1}{r}{$-1.3\cdot10^{-03}$} \\
& LightGCN & 0.0419 & 0.0323 & 0.0411 & 0.0315 & \multicolumn{1}{r}{$+8\cdot10^{-04}$} & \multicolumn{1}{r}{$+8\cdot10^{-04}$} \\
& SGL & 0.0474 & 0.0372 & 0.0478 & 0.0379 & \multicolumn{1}{r}{$-4\cdot10^{-04}$} & \multicolumn{1}{r}{$-7\cdot10^{-04}$} \\
& UltraGCN & 0.0688 & 0.0561 & 0.0681 & 0.0556 & \multicolumn{1}{r}{$+7\cdot10^{-04}$} & \multicolumn{1}{r}{$+5\cdot10^{-04}$} \\
& GFCF & 0.0710 & 0.0584 & 0.0710 & 0.0584 & \multicolumn{1}{r}{0} & \multicolumn{1}{r}{0} \\
\bottomrule
\multicolumn{8}{l}{\textit{*Results are not provided since SGL was not originally trained and tested on Gowalla~\cite{DBLP:conf/sigir/WuWF0CLX21}.}}
\end{tabular}
\end{table*}

\section{Benchmarking graph CF approaches using alternative baselines (RQ2)}
\label{sec:other-baselines}

In line with recent reproducibility works (such as \cite{DBLP:journals/tois/DacremaBCJ21}) that evidenced certain problems regarding the choice and optimization of the baselines used for comparison, in this section we assess how graph CF approaches perform relatively to classical CF baselines.
Here, we specify first how the experiments are prepared  (Section~\ref{bench_set}), and the corresponding results are shown (Section~\ref{bench_res}).
\subsection{Settings}
\label{bench_set}

We expand our investigation by examining four \textit{classic} CF models to enhance the replicability analysis. Specifically, we select four models whose accuracy performance has rarely been compared with the \textit{graph-based} CF approaches replicated in this study. The decision to include \userknn, \itemknn, \pbeta, and \easer is purposeful. We refer to~\cite{DBLP:conf/recsys/DacremaCJ19} and (more recently) \cite{DBLP:conf/um/AnelliBNJP22}, which demonstrated the competitiveness of these baseline models compared to more recent approaches when a shared benchmark for comparison is employed among all involved methodologies. Furthermore, we also consider two unpersonalized approaches (i.e., MostPop and Random). The two models act as benchmarks to assess the effectiveness of customized methods compared to a user-agnostic solution.

For a fair comparison, the configuration delineated herein elucidates
how the four \textit{classic} CF models are tuned following the exact
same training/test splitting reported in~Section \ref{sec:prior-results} and the same experimental protocol. The only difference is that (for obvious reasons) we need to explore the hyper-parameters of each classic CF model introduced in the comparison. Similarly to what the authors do in the original graph CF works, we retain the 10\% of the training to generate a validation set, but decide to explore 20 distinct configurations for each model through the state-of-the-art Tree-structured Parzen Estimator (TPE) hyper-parameter search~\cite{DBLP:conf/nips/BergstraBBK11}. For every model, the final results correspond to the accuracy measure on the test set by setting the hyper-parameter configuration providing the best Recall@20 results on the validation set. 
\subsection{Results}
\label{bench_res}

Table \ref{tab:baselines} shows the results of the graph CF models (as previously replicated in~Table \ref{tab:reproducibility})
with the additional baselines.
%
First, it is worth noting that, even though none of these baselines gets the best results in any of the three datasets considered, they achieve the second-best performance in Yelp 2018 (refer to \pbeta with nDCG).

Second, none of the models in the reference family achieve competitive performance.
While this is expected for the Random algorithm, it is an indication that either none of these datasets evidence a strong popularity bias or (considering the way they were processed) such bias was removed.

Third, some of the classic CF approaches
\iffalse
(such as \pbeta and UserkNN in Gowalla, \pbeta and \easer in Yelp, and X and Y in Amazon Book \alejandro{TODO})
\else
(such as \pbeta and UserkNN in Gowalla, and \pbeta and \easer in Yelp 2018)
\fi
demonstrate better performance than some of the state-of-the-art graph CF methods, in particular, they perform better than NGCF, DGCF, and LightGCN.
This result is in line with recent experimental comparisons~\cite{DBLP:journals/tois/DacremaBCJ21,DBLP:conf/recsys/AnelliBNP21,DBLP:conf/cikm/AnelliDNSFMP22} where these baselines outperform other methods based on matrix factorization or neural networks.
Moreover, to some extent, the fact that some graph CF methods are outperformed should not be surprising, since, as shown in~Table \ref{tab:models}, none of these baselines were included in the original papers where the graph CF approaches were proposed.
\begin{table}[!t]
\centering
\caption{Graph-based CF solutions tested against unpersonalized (i.e., reference) and classical CF approaches on Gowalla, Yelp 2018, and Amazon Book. While results for the graph-based approaches have been directly reported from our reproducibility study (see above), classical CF recommender systems have been fine-tuned on the two datasets to find their best configurations. \textbf{Boldface} and underline refer to best and second-to-best values, respectively.}
\label{tab:baselines}
\setlength{\tabcolsep}{3.5pt}
\footnotesize
\begin{tabular}{llcccccc}
\toprule
\textbf{Families} & \textbf{Models} & \multicolumn{2}{c}{\textbf{Gowalla}} & \multicolumn{2}{c}{\textbf{Yelp 2018}} & \multicolumn{2}{c}{\textbf{Amazon Book}} \\ \cmidrule{3-8} 
\multicolumn{2}{c}{} & \multicolumn{1}{c}{Recall} & \multicolumn{1}{c}{nDCG} & \multicolumn{1}{c}{Recall} & \multicolumn{1}{c}{nDCG} & \multicolumn{1}{c}{Recall} & \multicolumn{1}{c}{nDCG} \\ \cmidrule{1-8} 
\multirow{2}{*}{\textit{Reference}} & MostPop & 0.0416 & 0.0316 & 0.0125 & 0.0101 & 0.0051 & 0.0044 \\
& Random & 0.0005 & 0.0003 & 0.0005 & 0.0004 & 0.0002 & 0.0002 \\
\cmidrule{1-8} 
\multirow{4}{*}{\textit{Classic CF}} & \userknn & 0.1685 & 0.1370 & 0.0630 & 0.0528 & 0.0582 & 0.0477 \\
& \itemknn & 0.1409 & 0.1165 & 0.0610 & 0.0507 & 0.0634 & 0.0524 \\
& \pbeta & 0.1829 & 0.1520 & 0.0671 & 0.0559 & 0.0683 & 0.0565 \\
& \easer* & 0.1661 & 0.1384 & 0.0655 & 0.0552 & \textbf{0.0710} & \underline{0.0567} \\
\cmidrule{1-8} 
\multirow{6}{*}{\textit{Graph CF}} & NGCF & 0.1556 & 0.1320 & 0.0556 & 0.0452 & 0.0319 & 0.0246 \\
& DGCF & 0.1736 & 0.1477 & 0.0621 & 0.0505 & 0.0384 & 0.0295 \\
& LightGCN & 0.1826 & \underline{0.1545} & 0.0629 & 0.0516 & 0.0419 & 0.0323 \\
& SGL & \multicolumn{1}{c}{---} & --- & 0.0669 & 0.0552 & 0.0474 & 0.0372 \\
& UltraGCN & \textbf{0.1863} & \textbf{0.1580} & \underline{0.0672} & 0.0553 & \underline{0.0688} & 0.0561 \\
& GFCF & \underline{0.1849} & 0.1518 & \textbf{0.0697} & \textbf{0.0571} & \textbf{0.0710} & \textbf{0.0584} \\
\bottomrule
\multicolumn{8}{l}{\tiny \textit{*Results for \easer on Amazon Book are taken from \href{https://openbenchmark.github.io/BarsMatch/leaderboard/amazonbooks_m1.html}{\underline{BARS Benchmark}}~\cite{DBLP:conf/sigir/ZhuDSMLCXZ22}.}}
\end{tabular}
\end{table}


\section{Extending the experimental comparison to new datasets (RQ3 --- RQ4)}
\label{sec:other-datasets}

This section aims to provide a full picture from an experimental point of view on two new datasets: Allrecipes and BookCrossing.
First, Section~\ref{other_set} introduces the experimental settings followed to obtain the results presented in Section~\ref{other_res}. Then, Section~\ref{other_dis} discusses these results in more detail, aiming to explain the insights derived from them.

\subsection{Settings}
\label{other_set}

Motivated by the previous results,
we further enrich our analysis by investigating the behavior of all tested models on two datasets that have never been considered in any previous study involving graph-based approaches for recommendation, namely, Allrecipes~\cite{DBLP:journals/tmm/GaoFHHGFMC20} and BookCrossing~\cite{DBLP:conf/www/ZieglerMKL05}.
~Table \ref{tab:stats} shows some statistics of these datasets, where we purposely decide to report both the benchmarking datasets for graph CF (i.e., Gowalla, Yelp 2018, and Amazon Book) and the newly introduced ones. On the one hand, Allrecipes exhibits quite discordant characteristics compared to the other datasets. Although it has a comparable density, users are more numerous than items, with a much lower average user and item node degrees compared to the other standard graph CF datasets. On the other hand, BookCrossing displays the lowest ratio between the number of users to items across all datasets, and a much higher density than all the others. In summary, the newly introduced datasets serve as a foundation to assess the performance in different (and never-explored) \textit{topological} settings for graph CF baselines.

To adhere to the experimental setup presented so far, we adopt 
the all-unrated-item evaluation protocol, and split the two datasets with a random hold-out solution, ensuring an 80:20 proportion.
Differently from the replicability study, we now perform a TPE-based hyper-parameter tuning for \textit{all} models, as the best hyper-parameters for each graph-based approach is not known in advance; for this, we (again) use the 10\% portion of the training set as validation set.
We run 20 different settings within the search space provided in the original papers. The models' best configurations are selected through the Recall@20 on the validation.

\subsection{Results}
\label{other_res}

Table~\ref{tab:new_res} provides a full comparison between unpersonalized methods, classical CF approaches, and the graph CF methods under analysis.
In line with our previous section experiments, classic CF methods (in particular, \pbeta and \easer) are very competitive compared to graph CF approaches, even in novel datasets like the ones included in this analysis.
Specifically, the results in BookCrossing are dominated by these baselines,
while in Allrecipes, the MostPop stands out, evidencing a strong popularity bias.

These results highlight that, among the graph CF techniques, those that maintain their performance in novel domains are UltraGCN (best one in Allrecipes and third among its type) and LightGCN (second best in both domains). While the nature of these two datasets is clearly different (as shown in Table~\ref{tab:datasets}, Allrecipes is smaller and it contains more users than items, instead of the other way around as in BookCrossing), the relative performance of the best graph CF methods is competitive. However, for some of them, the performance drop is significant, reaching an accuracy lower than that of any other classic CF baseline.

To bring light into some of these behaviors, the next section discusses in more detail how the ranking of the graph CF methods changes depending on the dataset, and hypothesize which dataset characteristics may be tied to these effects.

\begin{table}[!t]
\centering
\caption{Statistics calculated on the training sets of Gowalla, Yelp 2018, Amazon Book, Allrecipes, and BookCrossing. We indicate the number of user-item interactions through' Edges' while 'Avg. Deg. (\textit{U})' and 'Avg. Deg. (\textit{I})' refer to users' and items' average node degree (i.e., average interaction number).}
\label{tab:stats}
\footnotesize
\setlength{\tabcolsep}{1.5pt}
\begin{tabular}{lrrrrr}
\toprule
\textbf{Statistics} & \textbf{Gowalla} & \textbf{Yelp 2018} & \textbf{Amazon Book} & \textbf{Allrecipes} & \textbf{BookCrossing} \\
\cmidrule{1-6}
Users & 29,858 & 31,668 & 52,643 & 10,084 & 6,754 \\
Items & 40,981 & 38,048 & 91,599 & 8,407 & 13,670 \\
Edges & 810,128 & 1,237,259 & 2,380,730 & 80,540 & 234,762 \\
Density & 0.0007 & 0.0010 & 0.0005 & 0.0010 & 0.0025 \\
Avg. Deg. (\textit{U}) & 27.1327 & 39.0697 & 45.2241 & 7.9869 & 34.7590 \\
Avg. Deg. (\textit{I}) & 19.7684 & 32.5184 & 25.9908 & 9.5801 & 17.1735 \\
\bottomrule
\end{tabular}
\end{table}

\begin{table}[!t]
\centering
\caption{Graph-based CF solutions tested against unpersonalized (i.e., reference) and classical CF approaches on Allrecipes and BookCrossing. Boldface and underline refer to best and second-to-best values, respectively.}
\label{tab:new_res}
\footnotesize
\begin{tabular}{lllcccc}
\toprule
\textbf{Families} & \textbf{Models} & \multicolumn{2}{c}{\textbf{Allrecipes}} & \multicolumn{2}{c}{\textbf{BookCrossing}} \\ \cmidrule{3-6} 
\multicolumn{2}{c}{} & \multicolumn{1}{c}{Recall} & \multicolumn{1}{c}{nDCG} & \multicolumn{1}{c}{Recall} & \multicolumn{1}{c}{nDCG} \\ \cmidrule{1-6} 
\multirow{2}{*}{\textit{Reference}} & MostPop & \underline{0.0472} & \underline{0.0242} & 0.0352 & 0.0319 \\
& Random & 0.0024 & 0.0010 & 0.0013 & 0.0011 \\
\cmidrule{1-6} 
\multirow{4}{*}{\textit{Classic CF}} & \userknn & 0.0339 & 0.0188 & 0.0871 & 0.0769 \\
& \itemknn & 0.0326 & 0.0180 & 0.0779 & 0.0739 \\
& \pbeta & 0.0170 & 0.0089 & \textbf{0.0941} & \underline{0.0821} \\
& \easer & 0.0351 & 0.0192 & \underline{0.0925} & \textbf{0.0847} \\
\cmidrule{1-6}
\multirow{6}{*}{\textit{Graph CF}} & NGCF & 0.0291 & 0.0144 & 0.0670 & 0.0546 \\
& DGCF & 0.0448 & 0.0234 & 0.0643 & 0.0543 \\
& LightGCN & 0.0459 & 0.0236 & 0.0803 & 0.0660 \\
& SGL & 0.0365 & 0.0192 & 0.0716 & 0.0600 \\
& UltraGCN & \textbf{0.0475} & \textbf{0.0248} & 0.0800 & 0.0651 \\
& GFCF & 0.0101 & 0.0051 & 0.0819 & 0.0712 \\
\bottomrule
\end{tabular}
\end{table}

\subsection{Discussion}
\label{other_dis}
To further validate and explain the reasons behind the results reported in~Table \ref{tab:new_res}, in the following we perform a twofold analysis. First, we rank all the selected graph-based recommendation models on all the tested datasets to assess their relative improvement across all settings and provide another perspective on the results from~Table \ref{tab:new_res}. Then, we propose a more nuanced study on the measured accuracy performance by investigating its (possible) dependence on the specific dataset characteristics, namely, the node degree as viewed at multiple hops. 

\subsubsection{Graph-based models' ranking} In~Table \ref{tab:rank}, we rank the six graph CF recommender systems under analysis according to the calculated Recall@20 and nDCG@20, for both the original datasets (i.e., Gowalla, Yelp 2018, and Amazon Book) and the novel datasets we introduced (i.e., Allrecipes and BookCrossing). Moreover, we also indicate the relative improvement of each model with respect to the worst-performing algorithm on that dataset. 

The trend on the three original datasets is quite steady, with UltraGCN and GFCF being the two best-performing approaches in almost all cases, and the remaining graph techniques ranked as in descending chronological order (confirming the findings from the recent literature). In terms of relative improvements, we observe large performance differences mainly on the Amazon Book setting.

By focusing on the two additional datasets (i.e., Allrecipes and BookCrossing), the rankings corroborate some of the previous outcomes, but also introduce novel and unexpected considerations. While UltraGCN seems to preserve its role of leading approach in the two scenarios (in BookCrossing it is ranked as third but with minimum margin to the second one), we notice how GFCF's performance is very fluctuating, as it even stands in the last position on Allrecipes with large performance difference to the other models (the same goes for DGCF). Noticeably, LightGCN gets up to the top of the ranking in both settings, indicating that a careful hyper-parameter tuning could be beneficial to outperform most of the other approaches, even the ones that should surpass it according to the literature (such as SGL). As final remarks, NGCF poor performance is again confirmed in such different dataset settings.

\subsubsection{Analysis on the node degree} As already observed in~Table \ref{tab:stats}, the average node degree of users and items represents one of the main aspects discerning each dataset from the other ones. For this reason, we decide to reason about its possible influence on the models' performance. In this respect, instead of limiting our analysis to the sole definition of node degree (i.e., number of recorded interactions for each user and item), and given the ability of graph-based approaches to distill the collaborative signal by stacking multiple layers~\cite{DBLP:conf/sigir/Wang0WFC19}, we propose a novel investigation which reinterprets the node degree as \textit{information flow} from neighbor nodes to the user nodes after multiple hops. Note that we only consider users as the ending nodes of such a flow because we are interested in assessing how the accuracy recommendation measures (which are generally calculated user-wise) may be influenced by this aspect.

\begin{table*}[!t]
\centering
\caption{Graph-based recommender systems, ranked according to their Recall@20 and nDCG@20 on all the tested datasets. For each model, we also report its relative improvement with respect to the worst-performing approach on the same dataset (in \textcolor{forestgreen(web)}{green}).}
\label{tab:rank}
\footnotesize
\begin{tabular}{lclllll}
\toprule
 \textbf{Metric} & \multicolumn{1}{c}{} & \textbf{Gowalla} & \textbf{Yelp 2018} & \textbf{Amazon Book} & \textbf{Allrecipes} & \textbf{BookCrossing} \\
\cmidrule{1-7}
\multirow{6}{*}{Recall} & \textbf{1.} & UltraGCN \textcolor{forestgreen(web)}{(\textit{+19.73\%})} & GFCF \textcolor{forestgreen(web)}{(\textit{+25.36\%})} & GFCF \textcolor{forestgreen(web)}{(\textit{+122.57\%})} & UltraGCN \textcolor{forestgreen(web)}{(\textit{+370.30\%})} & GFCF \textcolor{forestgreen(web)}{(\textit{+27.37\%})} \\
& \textbf{2.} & GFCF \textcolor{forestgreen(web)}{(\textit{+18.83\%})} & UltraGCN \textcolor{forestgreen(web)}{(\textit{+20.86\%})} & UltraGCN \textcolor{forestgreen(web)}{(\textit{+115.67\%})} & LightGCN \textcolor{forestgreen(web)}{(\textit{+354.46\%})} & LightGCN \textcolor{forestgreen(web)}{(\textit{+24.88\%})} \\
& \textbf{3.} & LightGCN \textcolor{forestgreen(web)}{(\textit{+17.35\%})} & SGL \textcolor{forestgreen(web)}{(\textit{+20.32\%})} & SGL \textcolor{forestgreen(web)}{(\textit{+48.59\%})} & DGCF \textcolor{forestgreen(web)}{(\textit{+343.56\%})} & UltraGCN \textcolor{forestgreen(web)}{(\textit{+24.42\%})} \\
& \textbf{4.} & DGCF \textcolor{forestgreen(web)}{(\textit{+11.57\%})} & LightGCN \textcolor{forestgreen(web)}{(\textit{+13.13\%})} & LightGCN \textcolor{forestgreen(web)}{(\textit{+31.35\%})} & SGL \textcolor{forestgreen(web)}{(\textit{+261.39\%})} & SGL \textcolor{forestgreen(web)}{(\textit{+11.35\%})} \\
& \textbf{5.} & NGCF ( --- ) & DGCF \textcolor{forestgreen(web)}{(\textit{+11.69\%})} & DGCF \textcolor{forestgreen(web)}{(\textit{+20.38\%})} &  NGCF \textcolor{forestgreen(web)}{(\textit{+188.12\%})} & NGCF \textcolor{forestgreen(web)}{(\textit{+4.20\%})} \\
& \textbf{6.} & \textit{SGL*} ( --- ) & NGCF ( --- ) & NGCF ( --- ) & GFCF ( --- ) & DGCF ( --- ) \\
\cmidrule{1-7}
\multirow{6}{*}{nDCG} & \textbf{1.} & UltraGCN \textcolor{forestgreen(web)}{(\textit{+19.70\%})} & GFCF \textcolor{forestgreen(web)}{(\textit{+26.33\%})} & GFCF \textcolor{forestgreen(web)}{(\textit{+137.40\%})} & UltraGCN \textcolor{forestgreen(web)}{(\textit{+386.27\%})} & GFCF \textcolor{forestgreen(web)}{(\textit{+31.12\%})} \\
& \textbf{2.} & LightGCN \textcolor{forestgreen(web)}{(\textit{+17.05\%})} & UltraGCN \textcolor{forestgreen(web)}{(\textit{+22.35\%})} & UltraGCN \textcolor{forestgreen(web)}{(\textit{+128.05\%})} & LightGCN \textcolor{forestgreen(web)}{(\textit{+362.75\%})} & LightGCN \textcolor{forestgreen(web)}{(\textit{+21.55\%})} \\
& \textbf{3.} & GFCF \textcolor{forestgreen(web)}{(\textit{+15.00\%})} & SGL \textcolor{forestgreen(web)}{(\textit{+22.12\%})} & SGL \textcolor{forestgreen(web)}{(\textit{+51.22\%})} & DGCF \textcolor{forestgreen(web)}{(\textit{+358.82\%})} & UltraGCN \textcolor{forestgreen(web)}{(\textit{+19.89\%})} \\
& \textbf{4.} & DGCF \textcolor{forestgreen(web)}{(\textit{+11.89\%})} & LightGCN \textcolor{forestgreen(web)}{(\textit{+14.16\%})} & LightGCN \textcolor{forestgreen(web)}{(\textit{+31.30\%})} & SGL \textcolor{forestgreen(web)}{(\textit{+276.47\%})} & SGL \textcolor{forestgreen(web)}{(\textit{+10.50\%})} \\
& \textbf{5.} & NGCF ( --- ) & DGCF \textcolor{forestgreen(web)}{(\textit{+11.73\%})} & DGCF \textcolor{forestgreen(web)}{(\textit{+19.92\%})} &  NGCF \textcolor{forestgreen(web)}{(\textit{+182.35\%})} & NGCF \textcolor{forestgreen(web)}{(\textit{+0.55\%})} \\
& \textbf{6.} & \textit{SGL*} ( --- ) & NGCF ( --- ) & NGCF ( --- ) & GFCF ( --- ) & DGCF ( --- ) \\
\bottomrule
\multicolumn{7}{l}{\textit{*SGL is not classifiable on the Gowalla dataset as results were not calculated in the original paper~\cite{DBLP:conf/sigir/WuWF0CLX21}.}}
\end{tabular}
\end{table*}

Before diving into the results and discussion, we provide some useful intuitions and formulations which may help understand our analysis. With reference to~Figure \ref{fig:hops}, we introduce the definition of information flow at one, two, and three hops. We decide to limit our focus on the first three explored hops because (i) graph-based recommender systems built upon the message-passing schema usually tend not to iterate over the third aggregation layer, and (ii) the investigation of more than three hops would not be meaningful from a recommendation perspective. As a matter of fact, we interpret each of the three hops as follows:
\begin{itemize}[leftmargin=*]
    \item at \textbf{one} hop (Figure \ref{fig:hops}a), users receive the information coming from the items they interacted with; in other words, this is an indication of the \textit{activeness} of users on the platform;
    \item at \textbf{two} hops (Figure \ref{fig:hops}b), users receive the information of the other users co-interacting with the same items; in other words, this is an indication of the influence of \textit{items' popularity} on users;
    \item at \textbf{three} hops (Figure \ref{fig:hops}c), users receive the information coming from the items interacted by the other users involved in co-interactions; 
    i.e., 
    this is an indication of the influence of \textit{co-interacting} users' \textit{activeness} on users.
\end{itemize}

Let us formalize such definitions. The information received by users at one, two, and three hops is calculated as:
\begin{equation}
\small
    \boldsymbol{\Upsilon}_\mathcal{U}^{(1)} = \mathbf{R} \mathbf{1}_\mathcal{I}, \qquad \boldsymbol{\Upsilon}_\mathcal{U}^{(2)} = (\mathbf{R} \odot (\mathbf{1}_{\mathcal{U}} \mathbf{R})) \mathbf{1}_{\mathcal{I}}, \qquad \boldsymbol{\Upsilon}_\mathcal{U}^{(3)} = (\mathbf{R}\mathbf{R}^{\top} \odot \mathbf{R}\mathbf{1}_{\mathcal{I}})\mathbf{1}_{\mathcal{I}},
\end{equation}
where $\boldsymbol{\Upsilon}_\mathcal{U}^{(h)} \in \mathbb{R}^{|\mathcal{U}| \times 1}$ is the vector of the information that all users receive from the nodes in their $h$-hop, $\mathbf{1}_\mathcal{U} \in \mathbb{R}^{1 \times |\mathcal{U}|}$ and $\mathbf{1}_\mathcal{I} \in \mathbb{R}^{|\mathcal{I}| \times 1}$ are row and column vectors with 1 repeated $|\mathcal{U}|$ and $|\mathcal{I}|$ times, respectively, while $\odot$ is the Hadamard product performed in broadcast. 

\input{figures/hops}

In light of the above, the study assesses the accuracy performance of graph-based recommender systems on user groups considering the information received from the one, two, and three hops neighborhood. Following other analyses in the literature, we decide to split users into quartiles according to the information values (i.e., $\boldsymbol{\Upsilon}^{(h)}_\mathcal{U}$). Thus, we consider four groups: (i) users whose values are below the 25\% of the distribution, (ii) users whose values are above the 25\% and below the 50\% of the distribution, (iii) users whose values are above the 50\% and below the 75\% of the distribution, and (iv) users whose values are above the 75\% of the distribution.


Figure \ref{fig:layers_perc} displays the percentage variation in accuracy performance (measured by nDCG) across quartiles relative to the average value reported in~Table \ref{tab:new_res}. The figure illustrates how the quality of recommendation performance fluctuates amongst different clusters of users.
For example, a method indicating a 50\% improvement in the fourth quartile would suggest that users in this cluster, typically more active (1-hop) or also interested in popular items (2-hop), receive more accurate recommendations with respect to the average user. 
This observation implies that a non-discriminatory recommendation system should produce no variation across quartiles, with values overlapping the 0\% dashed line. 
The second necessary preliminary to understand the outcome of the experiments is the interpretation of the quartiles for the different hops.
In the 1-hop, the fourth quartile pertains to warm users interacting most with the platform, while the first quartile represents cold users interacting less frequently. In the 2-hop, high values in the fourth quartile indicate active users who enjoy popular items, resulting in dense subgraphs. The first quartile, in contrast, consists of less active users interacting with niche items in less dense subgraphs. The 3-hop, which includes user neighbors, generates the highest values when active users interact with popular items enjoyed by warm users (i.e., their neighbors).
However, it is essential to note that the plots offer no insight into overall accuracy (which is in~Table \ref{tab:new_res}).

\input{figures/quartiles_all_charts}

When considering the recommendation performance according to the corresponding cluster (depicted in~Figure \ref{fig:layers_perc}), it is crucial to note that none of them demonstrate ideal recommendation behavior. Instead, these systems tend to favor warm users or densely interconnected subgraphs located in the fourth quartile.
Despite this trend, the 1-hop plots for graph Collaborative Filtering (CF) and classic CF methods in Allrecipes and BookCrossing graphs demonstrate minimal disparities between different recommendation approaches. Even though they all favor the fourth quartile over the first one, the coldness/warmness of a user marginally impacts how much the method is biased toward these types of users. The lone exception to this trend is GFCF, which exhibits even greater penalization towards the first three quartiles (varying on the three hops from, approximately, -45\% to +115\%, and thus exceeding the plots' upper bound). As such, this system only provides satisfactory recommendation performance for users in the fourth quartile.

Regarding the 2-hop, there are several interesting insights to be gained. Firstly, the recommendation methods exhibit a higher overall slope, favoring the users who enjoyed popular items over the cold users who enjoyed niche items. While this may seem like an obvious observation, the plot confirms that user coldness/warmness alone is not a sufficient indicator of high-quality recommendations. 
Instead, the 2-hop reveals that combining user coldness/warmness and item popularity is useful for identifying such users.
A second noteworthy aspect is that the Allrecipes dataset highlights three distinct behaviors among the graph CF methods. UltraGCN, DGCF, and LightGCN exhibit similar performance and display less discriminatory behavior across quartiles. It is interesting to note that these models also perform best overall (see~Table \ref{tab:new_res}). On the other hand, SGL and NGCF show a higher slope that is comparable to classic 
CF methods.
Also, their corresponding performance is similar in~Table \ref{tab:new_res}.
A third observation concerns GFCF, which performs poorly across all quartiles except for the fourth. Its behavior is even more accentuated than in the 1-hop analysis. Additionally, NGCF, SGL, and GFCF are graph CF algorithms performing differently according to user warmness and item popularity. 
Meanwhile, all algorithms in BookCrossing, and the classic CF in Allrecipes, exhibit the distribution over the quartiles across methods.

Finally, in the 3-hop, for the BookCrossing dataset, the information pertaining to neighbors does not contribute significantly to the results, as indicated by the similarity between the 2- and 3-hop plots. 
Meanwhile, in Allrecipes, the best models (UltraGCN, DGCF, and LightGCN) exhibit more consistency in performance across all quartiles, as demonstrated by a more even distribution of results (less variations across the quartiles). However, this pattern is not evident in NGCF, SGL, and GFCF, which exhibit a more disparate range of results across the quartiles.

\section{Conclusion and Future Work}
\label{sec:conclusion}

This study replicates the results of six graph CF methods, namely NGCF, DGCF, LightGCN, SGL, UltraGCN, and GFCF, and expands the research to include state-of-the-art recommendation strategies like \userknn, \itemknn, \pbeta, and \easer. The observed high rankings of the latter ones highlight the need for more comprehensive evaluations. After the initial study on the standard Gowalla, Yelp 2018, and Amazon Book datasets, experiments are extended to two additional datasets, Allrecipes and BookCrossing, which reveal substantial ranking variations compared to the initial datasets. Thus, the study introduces and analyzes the information flow in the graph and discovers that 2-hop information (combining user activeness and item popularity) is a valid indicator of CF behavior and could motivate the recommendation performance. The experimental results call for further investigations into the diversity and fairness of the considered methods and whether graph methods effectively mitigate popularity bias.


\begin{acks}
This work was partially supported by the following projects: Secure Safe Apulia,
MISE CUP: I14E20000020001 CTEMT - Casa delle Tecnologie Emergenti Comune di Matera, CT\_FINCONS\_II,  CT\_FINCONS\_III, OVS Fashion Retail Reloaded, LUTECH DIGITALE 4.0, PID2019-108965GB-I00.
\end{acks}

\bibliographystyle{ACM-Reference-Format}
\bibliography{bibliography}


\begin{thebibliography}{97}


\ifx \showCODEN    \undefined \def \showCODEN     #1{\unskip}     \fi
\ifx \showDOI      \undefined \def \showDOI       #1{#1}\fi
\ifx \showISBNx    \undefined \def \showISBNx     #1{\unskip}     \fi
\ifx \showISBNxiii \undefined \def \showISBNxiii  #1{\unskip}     \fi
\ifx \showISSN     \undefined \def \showISSN      #1{\unskip}     \fi
\ifx \showLCCN     \undefined \def \showLCCN      #1{\unskip}     \fi
\ifx \shownote     \undefined \def \shownote      #1{#1}          \fi
\ifx \showarticletitle \undefined \def \showarticletitle #1{#1}   \fi
\ifx \showURL      \undefined \def \showURL       {\relax}        \fi
\providecommand\bibfield[2]{#2}
\providecommand\bibinfo[2]{#2}
\providecommand\natexlab[1]{#1}
\providecommand\showeprint[2][]{arXiv:#2}

\bibitem[Anelli et~al\mbox{.}(2021a)]%
        {DBLP:conf/sigir/AnelliBFMMPDN21}
\bibfield{author}{\bibinfo{person}{Vito~Walter Anelli},
  \bibinfo{person}{Alejandro Bellog{\'{\i}}n}, \bibinfo{person}{Antonio
  Ferrara}, \bibinfo{person}{Daniele Malitesta},
  \bibinfo{person}{Felice~Antonio Merra}, \bibinfo{person}{Claudio Pomo},
  \bibinfo{person}{Francesco~Maria Donini}, {and} \bibinfo{person}{Tommaso~Di
  Noia}.} \bibinfo{year}{2021}\natexlab{a}.
\newblock \showarticletitle{Elliot: {A} Comprehensive and Rigorous Framework
  for Reproducible Recommender Systems Evaluation}. In
  \bibinfo{booktitle}{\emph{{SIGIR}}}. \bibinfo{publisher}{{ACM}},
  \bibinfo{pages}{2405--2414}.
\newblock


\bibitem[Anelli et~al\mbox{.}(2022a)]%
        {DBLP:conf/um/AnelliBNJP22}
\bibfield{author}{\bibinfo{person}{Vito~Walter Anelli},
  \bibinfo{person}{Alejandro Bellog{\'{\i}}n}, \bibinfo{person}{Tommaso~Di
  Noia}, \bibinfo{person}{Dietmar Jannach}, {and} \bibinfo{person}{Claudio
  Pomo}.} \bibinfo{year}{2022}\natexlab{a}.
\newblock \showarticletitle{Top-N Recommendation Algorithms: {A} Quest for the
  State-of-the-Art}. In \bibinfo{booktitle}{\emph{{UMAP}}}.
  \bibinfo{publisher}{{ACM}}, \bibinfo{pages}{121--131}.
\newblock


\bibitem[Anelli et~al\mbox{.}(2021b)]%
        {DBLP:conf/recsys/AnelliBNP21}
\bibfield{author}{\bibinfo{person}{Vito~Walter Anelli},
  \bibinfo{person}{Alejandro Bellog{\'{\i}}n}, \bibinfo{person}{Tommaso~Di
  Noia}, {and} \bibinfo{person}{Claudio Pomo}.}
  \bibinfo{year}{2021}\natexlab{b}.
\newblock \showarticletitle{Reenvisioning the comparison between Neural
  Collaborative Filtering and Matrix Factorization}. In
  \bibinfo{booktitle}{\emph{RecSys}}. \bibinfo{publisher}{{ACM}},
  \bibinfo{pages}{521--529}.
\newblock


\bibitem[Anelli et~al\mbox{.}(2023)]%
        {DBLP:conf/ecir/AnelliDNMPP23}
\bibfield{author}{\bibinfo{person}{Vito~Walter Anelli}, \bibinfo{person}{Yashar
  Deldjoo}, \bibinfo{person}{Tommaso~Di Noia}, \bibinfo{person}{Daniele
  Malitesta}, \bibinfo{person}{Vincenzo Paparella}, {and}
  \bibinfo{person}{Claudio Pomo}.} \bibinfo{year}{2023}\natexlab{}.
\newblock \showarticletitle{Auditing Consumer- and Producer-Fairness in Graph
  Collaborative Filtering}. In \bibinfo{booktitle}{\emph{{ECIR} {(1)}}}
  \emph{(\bibinfo{series}{Lecture Notes in Computer Science},
  Vol.~\bibinfo{volume}{13980})}. \bibinfo{publisher}{Springer},
  \bibinfo{pages}{33--48}.
\newblock


\bibitem[Anelli et~al\mbox{.}(2022b)]%
        {DBLP:conf/recsys/AnelliDNSFMP22}
\bibfield{author}{\bibinfo{person}{Vito~Walter Anelli}, \bibinfo{person}{Yashar
  Deldjoo}, \bibinfo{person}{Tommaso~Di Noia}, \bibinfo{person}{Eugenio~Di
  Sciascio}, \bibinfo{person}{Antonio Ferrara}, \bibinfo{person}{Daniele
  Malitesta}, {and} \bibinfo{person}{Claudio Pomo}.}
  \bibinfo{year}{2022}\natexlab{b}.
\newblock \showarticletitle{How Neighborhood Exploration influences Novelty and
  Diversity in Graph Collaborative Filtering}. In
  \bibinfo{booktitle}{\emph{MORS@RecSys}} \emph{(\bibinfo{series}{{CEUR}
  Workshop Proceedings}, Vol.~\bibinfo{volume}{3268})}.
  \bibinfo{publisher}{CEUR-WS.org}.
\newblock


\bibitem[Anelli et~al\mbox{.}(2022c)]%
        {DBLP:conf/cikm/AnelliDNSFMP22}
\bibfield{author}{\bibinfo{person}{Vito~Walter Anelli}, \bibinfo{person}{Yashar
  Deldjoo}, \bibinfo{person}{Tommaso~Di Noia}, \bibinfo{person}{Eugenio~Di
  Sciascio}, \bibinfo{person}{Antonio Ferrara}, \bibinfo{person}{Daniele
  Malitesta}, {and} \bibinfo{person}{Claudio Pomo}.}
  \bibinfo{year}{2022}\natexlab{c}.
\newblock \showarticletitle{Reshaping Graph Recommendation with Edge Graph
  Collaborative Filtering and Customer Reviews}. In
  \bibinfo{booktitle}{\emph{DL4SR@CIKM}} \emph{(\bibinfo{series}{{CEUR}
  Workshop Proceedings}, Vol.~\bibinfo{volume}{3317})}.
  \bibinfo{publisher}{CEUR-WS.org}.
\newblock


\bibitem[Bellog{\'{\i}}n and Said(2021)]%
        {DBLP:journals/umuai/BelloginS21}
\bibfield{author}{\bibinfo{person}{Alejandro Bellog{\'{\i}}n} {and}
  \bibinfo{person}{Alan Said}.} \bibinfo{year}{2021}\natexlab{}.
\newblock \showarticletitle{Improving accountability in recommender systems
  research through reproducibility}.
\newblock \bibinfo{journal}{\emph{User Model. User Adapt. Interact.}}
  \bibinfo{volume}{31}, \bibinfo{number}{5} (\bibinfo{year}{2021}),
  \bibinfo{pages}{941--977}.
\newblock


\bibitem[Bergstra et~al\mbox{.}(2011)]%
        {DBLP:conf/nips/BergstraBBK11}
\bibfield{author}{\bibinfo{person}{James Bergstra}, \bibinfo{person}{R{\'{e}}mi
  Bardenet}, \bibinfo{person}{Yoshua Bengio}, {and}
  \bibinfo{person}{Bal{\'{a}}zs K{\'{e}}gl}.} \bibinfo{year}{2011}\natexlab{}.
\newblock \showarticletitle{Algorithms for Hyper-Parameter Optimization}. In
  \bibinfo{booktitle}{\emph{{NIPS}}}. \bibinfo{pages}{2546--2554}.
\newblock


\bibitem[Bronstein et~al\mbox{.}(2021)]%
        {DBLP:journals/corr/abs-2104-13478}
\bibfield{author}{\bibinfo{person}{Michael~M. Bronstein}, \bibinfo{person}{Joan
  Bruna}, \bibinfo{person}{Taco Cohen}, {and} \bibinfo{person}{Petar
  Velickovic}.} \bibinfo{year}{2021}\natexlab{}.
\newblock \showarticletitle{Geometric Deep Learning: Grids, Groups, Graphs,
  Geodesics, and Gauges}.
\newblock \bibinfo{journal}{\emph{CoRR}}  \bibinfo{volume}{abs/2104.13478}
  (\bibinfo{year}{2021}).
\newblock


\bibitem[Cao et~al\mbox{.}(2021)]%
        {DBLP:conf/wsdm/CaoLGLL021}
\bibfield{author}{\bibinfo{person}{Jiangxia Cao}, \bibinfo{person}{Xixun Lin},
  \bibinfo{person}{Shu Guo}, \bibinfo{person}{Luchen Liu},
  \bibinfo{person}{Tingwen Liu}, {and} \bibinfo{person}{Bin Wang}.}
  \bibinfo{year}{2021}\natexlab{}.
\newblock \showarticletitle{Bipartite Graph Embedding via Mutual Information
  Maximization}. In \bibinfo{booktitle}{\emph{{WSDM}}}.
  \bibinfo{publisher}{{ACM}}, \bibinfo{pages}{635--643}.
\newblock


\bibitem[Chen et~al\mbox{.}(2020c)]%
        {DBLP:journals/tois/ChenZZLM20}
\bibfield{author}{\bibinfo{person}{Chong Chen}, \bibinfo{person}{Min Zhang},
  \bibinfo{person}{Yongfeng Zhang}, \bibinfo{person}{Yiqun Liu}, {and}
  \bibinfo{person}{Shaoping Ma}.} \bibinfo{year}{2020}\natexlab{c}.
\newblock \showarticletitle{Efficient Neural Matrix Factorization without
  Sampling for Recommendation}.
\newblock \bibinfo{journal}{\emph{{ACM} Trans. Inf. Syst.}}
  \bibinfo{volume}{38}, \bibinfo{number}{2} (\bibinfo{year}{2020}),
  \bibinfo{pages}{14:1--14:28}.
\newblock


\bibitem[Chen et~al\mbox{.}(2020a)]%
        {DBLP:conf/aaai/ChenLLLZS20}
\bibfield{author}{\bibinfo{person}{Deli Chen}, \bibinfo{person}{Yankai Lin},
  \bibinfo{person}{Wei Li}, \bibinfo{person}{Peng Li}, \bibinfo{person}{Jie
  Zhou}, {and} \bibinfo{person}{Xu Sun}.} \bibinfo{year}{2020}\natexlab{a}.
\newblock \showarticletitle{Measuring and Relieving the Over-Smoothing Problem
  for Graph Neural Networks from the Topological View}. In
  \bibinfo{booktitle}{\emph{{AAAI}}}. \bibinfo{publisher}{{AAAI} Press},
  \bibinfo{pages}{3438--3445}.
\newblock


\bibitem[Chen et~al\mbox{.}(2022)]%
        {DBLP:conf/wsdm/Chen0S0ZZ22}
\bibfield{author}{\bibinfo{person}{Hanxiong Chen}, \bibinfo{person}{Yunqi Li},
  \bibinfo{person}{Shaoyun Shi}, \bibinfo{person}{Shuchang Liu},
  \bibinfo{person}{He Zhu}, {and} \bibinfo{person}{Yongfeng Zhang}.}
  \bibinfo{year}{2022}\natexlab{}.
\newblock \showarticletitle{Graph Collaborative Reasoning}. In
  \bibinfo{booktitle}{\emph{{WSDM}}}. \bibinfo{publisher}{{ACM}},
  \bibinfo{pages}{75--84}.
\newblock


\bibitem[Chen et~al\mbox{.}(2020b)]%
        {DBLP:conf/aaai/ChenWHZW20}
\bibfield{author}{\bibinfo{person}{Lei Chen}, \bibinfo{person}{Le Wu},
  \bibinfo{person}{Richang Hong}, \bibinfo{person}{Kun Zhang}, {and}
  \bibinfo{person}{Meng Wang}.} \bibinfo{year}{2020}\natexlab{b}.
\newblock \showarticletitle{Revisiting Graph Based Collaborative Filtering: {A}
  Linear Residual Graph Convolutional Network Approach}. In
  \bibinfo{booktitle}{\emph{{AAAI}}}. \bibinfo{publisher}{{AAAI} Press},
  \bibinfo{pages}{27--34}.
\newblock


\bibitem[Dacrema et~al\mbox{.}(2021)]%
        {DBLP:journals/tois/DacremaBCJ21}
\bibfield{author}{\bibinfo{person}{Maurizio~Ferrari Dacrema},
  \bibinfo{person}{Simone Boglio}, \bibinfo{person}{Paolo Cremonesi}, {and}
  \bibinfo{person}{Dietmar Jannach}.} \bibinfo{year}{2021}\natexlab{}.
\newblock \showarticletitle{A Troubling Analysis of Reproducibility and
  Progress in Recommender Systems Research}.
\newblock \bibinfo{journal}{\emph{{ACM} Trans. Inf. Syst.}}
  \bibinfo{volume}{39}, \bibinfo{number}{2} (\bibinfo{year}{2021}),
  \bibinfo{pages}{20:1--20:49}.
\newblock


\bibitem[Dacrema et~al\mbox{.}(2019)]%
        {DBLP:conf/recsys/DacremaCJ19}
\bibfield{author}{\bibinfo{person}{Maurizio~Ferrari Dacrema},
  \bibinfo{person}{Paolo Cremonesi}, {and} \bibinfo{person}{Dietmar Jannach}.}
  \bibinfo{year}{2019}\natexlab{}.
\newblock \showarticletitle{Are we really making much progress? {A} worrying
  analysis of recent neural recommendation approaches}. In
  \bibinfo{booktitle}{\emph{RecSys}}. \bibinfo{publisher}{{ACM}},
  \bibinfo{pages}{101--109}.
\newblock


\bibitem[Du et~al\mbox{.}(2022)]%
        {DBLP:conf/mm/DuWF0022}
\bibfield{author}{\bibinfo{person}{Xiaoyu Du}, \bibinfo{person}{Zike Wu},
  \bibinfo{person}{Fuli Feng}, \bibinfo{person}{Xiangnan He}, {and}
  \bibinfo{person}{Jinhui Tang}.} \bibinfo{year}{2022}\natexlab{}.
\newblock \showarticletitle{Invariant Representation Learning for Multimedia
  Recommendation}. In \bibinfo{booktitle}{\emph{{ACM} Multimedia}}.
  \bibinfo{publisher}{{ACM}}, \bibinfo{pages}{619--628}.
\newblock


\bibitem[Ebesu et~al\mbox{.}(2018)]%
        {DBLP:conf/sigir/EbesuSF18}
\bibfield{author}{\bibinfo{person}{Travis Ebesu}, \bibinfo{person}{Bin Shen},
  {and} \bibinfo{person}{Yi Fang}.} \bibinfo{year}{2018}\natexlab{}.
\newblock \showarticletitle{Collaborative Memory Network for Recommendation
  Systems}. In \bibinfo{booktitle}{\emph{{SIGIR}}}. \bibinfo{publisher}{{ACM}},
  \bibinfo{pages}{515--524}.
\newblock


\bibitem[Fan et~al\mbox{.}(2022)]%
        {DBLP:conf/sigir/FanL0ZT022}
\bibfield{author}{\bibinfo{person}{Wenqi Fan}, \bibinfo{person}{Xiaorui Liu},
  \bibinfo{person}{Wei Jin}, \bibinfo{person}{Xiangyu Zhao},
  \bibinfo{person}{Jiliang Tang}, {and} \bibinfo{person}{Qing Li}.}
  \bibinfo{year}{2022}\natexlab{}.
\newblock \showarticletitle{Graph Trend Filtering Networks for Recommendation}.
  In \bibinfo{booktitle}{\emph{{SIGIR}}}. \bibinfo{publisher}{{ACM}},
  \bibinfo{pages}{112--121}.
\newblock


\bibitem[Gao et~al\mbox{.}(2022b)]%
        {DBLP:conf/wsdm/GaoW0022}
\bibfield{author}{\bibinfo{person}{Chen Gao}, \bibinfo{person}{Xiang Wang},
  \bibinfo{person}{Xiangnan He}, {and} \bibinfo{person}{Yong Li}.}
  \bibinfo{year}{2022}\natexlab{b}.
\newblock \showarticletitle{Graph Neural Networks for Recommender System}. In
  \bibinfo{booktitle}{\emph{{WSDM}}}. \bibinfo{publisher}{{ACM}},
  \bibinfo{pages}{1623--1625}.
\newblock


\bibitem[Gao et~al\mbox{.}(2020)]%
        {DBLP:journals/tmm/GaoFHHGFMC20}
\bibfield{author}{\bibinfo{person}{Xiaoyan Gao}, \bibinfo{person}{Fuli Feng},
  \bibinfo{person}{Xiangnan He}, \bibinfo{person}{Heyan Huang},
  \bibinfo{person}{Xinyu Guan}, \bibinfo{person}{Chong Feng},
  \bibinfo{person}{Zhaoyan Ming}, {and} \bibinfo{person}{Tat{-}Seng Chua}.}
  \bibinfo{year}{2020}\natexlab{}.
\newblock \showarticletitle{Hierarchical Attention Network for Visually-Aware
  Food Recommendation}.
\newblock \bibinfo{journal}{\emph{{IEEE} Trans. Multim.}} \bibinfo{volume}{22},
  \bibinfo{number}{6} (\bibinfo{year}{2020}), \bibinfo{pages}{1647--1659}.
\newblock


\bibitem[Gao et~al\mbox{.}(2022a)]%
        {DBLP:conf/sigir/Gao0HCZFZ22}
\bibfield{author}{\bibinfo{person}{Yunjun Gao}, \bibinfo{person}{Yuntao Du},
  \bibinfo{person}{Yujia Hu}, \bibinfo{person}{Lu Chen},
  \bibinfo{person}{Xinjun Zhu}, \bibinfo{person}{Ziquan Fang}, {and}
  \bibinfo{person}{Baihua Zheng}.} \bibinfo{year}{2022}\natexlab{a}.
\newblock \showarticletitle{Self-Guided Learning to Denoise for Robust
  Recommendation}. In \bibinfo{booktitle}{\emph{{SIGIR}}}.
  \bibinfo{publisher}{{ACM}}, \bibinfo{pages}{1412--1422}.
\newblock


\bibitem[Gilmer et~al\mbox{.}(2017)]%
        {DBLP:conf/icml/GilmerSRVD17}
\bibfield{author}{\bibinfo{person}{Justin Gilmer}, \bibinfo{person}{Samuel~S.
  Schoenholz}, \bibinfo{person}{Patrick~F. Riley}, \bibinfo{person}{Oriol
  Vinyals}, {and} \bibinfo{person}{George~E. Dahl}.}
  \bibinfo{year}{2017}\natexlab{}.
\newblock \showarticletitle{Neural Message Passing for Quantum Chemistry}. In
  \bibinfo{booktitle}{\emph{{ICML}}} \emph{(\bibinfo{series}{Proceedings of
  Machine Learning Research}, Vol.~\bibinfo{volume}{70})}.
  \bibinfo{publisher}{{PMLR}}, \bibinfo{pages}{1263--1272}.
\newblock


\bibitem[Gong et~al\mbox{.}(2022)]%
        {DBLP:conf/cikm/GongSWLL22}
\bibfield{author}{\bibinfo{person}{Kaiqi Gong}, \bibinfo{person}{Xiao Song},
  \bibinfo{person}{Senzhang Wang}, \bibinfo{person}{Songsong Liu}, {and}
  \bibinfo{person}{Yong Li}.} \bibinfo{year}{2022}\natexlab{}.
\newblock \showarticletitle{{ITSM-GCN:} Informative Training Sample Mining for
  Graph Convolutional Network-based Collaborative Filtering}. In
  \bibinfo{booktitle}{\emph{{CIKM}}}. \bibinfo{publisher}{{ACM}},
  \bibinfo{pages}{614--623}.
\newblock


\bibitem[Grover and Leskovec(2016)]%
        {DBLP:conf/kdd/GroverL16}
\bibfield{author}{\bibinfo{person}{Aditya Grover} {and} \bibinfo{person}{Jure
  Leskovec}.} \bibinfo{year}{2016}\natexlab{}.
\newblock \showarticletitle{node2vec: Scalable Feature Learning for Networks}.
  In \bibinfo{booktitle}{\emph{{KDD}}}. \bibinfo{publisher}{{ACM}},
  \bibinfo{pages}{855--864}.
\newblock


\bibitem[Hamilton(2020)]%
        {DBLP:series/synthesis/2020Hamilton}
\bibfield{author}{\bibinfo{person}{William~L. Hamilton}.}
  \bibinfo{year}{2020}\natexlab{}.
\newblock \bibinfo{booktitle}{\emph{Graph Representation Learning}}.
\newblock \bibinfo{publisher}{Morgan {\&} Claypool Publishers}.
\newblock


\bibitem[He and McAuley(2016)]%
        {DBLP:conf/www/HeM16}
\bibfield{author}{\bibinfo{person}{Ruining He} {and} \bibinfo{person}{Julian~J.
  McAuley}.} \bibinfo{year}{2016}\natexlab{}.
\newblock \showarticletitle{Ups and Downs: Modeling the Visual Evolution of
  Fashion Trends with One-Class Collaborative Filtering}. In
  \bibinfo{booktitle}{\emph{{WWW}}}. \bibinfo{publisher}{{ACM}},
  \bibinfo{pages}{507--517}.
\newblock


\bibitem[He et~al\mbox{.}(2020)]%
        {DBLP:conf/sigir/0001DWLZ020}
\bibfield{author}{\bibinfo{person}{Xiangnan He}, \bibinfo{person}{Kuan Deng},
  \bibinfo{person}{Xiang Wang}, \bibinfo{person}{Yan Li},
  \bibinfo{person}{Yong{-}Dong Zhang}, {and} \bibinfo{person}{Meng Wang}.}
  \bibinfo{year}{2020}\natexlab{}.
\newblock \showarticletitle{LightGCN: Simplifying and Powering Graph
  Convolution Network for Recommendation}. In
  \bibinfo{booktitle}{\emph{{SIGIR}}}. \bibinfo{publisher}{{ACM}},
  \bibinfo{pages}{639--648}.
\newblock


\bibitem[He et~al\mbox{.}(2017)]%
        {DBLP:conf/www/HeLZNHC17}
\bibfield{author}{\bibinfo{person}{Xiangnan He}, \bibinfo{person}{Lizi Liao},
  \bibinfo{person}{Hanwang Zhang}, \bibinfo{person}{Liqiang Nie},
  \bibinfo{person}{Xia Hu}, {and} \bibinfo{person}{Tat{-}Seng Chua}.}
  \bibinfo{year}{2017}\natexlab{}.
\newblock \showarticletitle{Neural Collaborative Filtering}. In
  \bibinfo{booktitle}{\emph{{WWW}}}. \bibinfo{publisher}{{ACM}},
  \bibinfo{pages}{173--182}.
\newblock


\bibitem[Hsieh et~al\mbox{.}(2017)]%
        {DBLP:conf/www/HsiehYCLBE17}
\bibfield{author}{\bibinfo{person}{Cheng{-}Kang Hsieh}, \bibinfo{person}{Longqi
  Yang}, \bibinfo{person}{Yin Cui}, \bibinfo{person}{Tsung{-}Yi Lin},
  \bibinfo{person}{Serge~J. Belongie}, {and} \bibinfo{person}{Deborah Estrin}.}
  \bibinfo{year}{2017}\natexlab{}.
\newblock \showarticletitle{Collaborative Metric Learning}. In
  \bibinfo{booktitle}{\emph{{WWW}}}. \bibinfo{publisher}{{ACM}},
  \bibinfo{pages}{193--201}.
\newblock


\bibitem[Huang et~al\mbox{.}(2022)]%
        {DBLP:conf/cikm/HuangXW0Y22}
\bibfield{author}{\bibinfo{person}{Chao Huang}, \bibinfo{person}{Lianghao Xia},
  \bibinfo{person}{Xiang Wang}, \bibinfo{person}{Xiangnan He}, {and}
  \bibinfo{person}{Dawei Yin}.} \bibinfo{year}{2022}\natexlab{}.
\newblock \showarticletitle{Self-Supervised Learning for Recommendation}. In
  \bibinfo{booktitle}{\emph{{CIKM}}}. \bibinfo{publisher}{{ACM}},
  \bibinfo{pages}{5136--5139}.
\newblock


\bibitem[Huang et~al\mbox{.}(2021)]%
        {DBLP:conf/kdd/HuangDDYFW021}
\bibfield{author}{\bibinfo{person}{Tinglin Huang}, \bibinfo{person}{Yuxiao
  Dong}, \bibinfo{person}{Ming Ding}, \bibinfo{person}{Zhen Yang},
  \bibinfo{person}{Wenzheng Feng}, \bibinfo{person}{Xinyu Wang}, {and}
  \bibinfo{person}{Jie Tang}.} \bibinfo{year}{2021}\natexlab{}.
\newblock \showarticletitle{MixGCF: An Improved Training Method for Graph
  Neural Network-based Recommender Systems}. In
  \bibinfo{booktitle}{\emph{{KDD}}}. \bibinfo{publisher}{{ACM}},
  \bibinfo{pages}{665--674}.
\newblock


\bibitem[Jordan(2023)]%
        {jordan2023calibrated}
\bibfield{author}{\bibinfo{person}{Keller Jordan}.}
  \bibinfo{year}{2023}\natexlab{}.
\newblock \bibinfo{title}{Calibrated Chaos: Variance Between Runs of Neural
  Network Training is Harmless and Inevitable}.
\newblock
\newblock
\showeprint[arxiv]{2304.01910}~[cs.LG]


\bibitem[Khosla et~al\mbox{.}(2020)]%
        {DBLP:conf/nips/KhoslaTWSTIMLK20}
\bibfield{author}{\bibinfo{person}{Prannay Khosla}, \bibinfo{person}{Piotr
  Teterwak}, \bibinfo{person}{Chen Wang}, \bibinfo{person}{Aaron Sarna},
  \bibinfo{person}{Yonglong Tian}, \bibinfo{person}{Phillip Isola},
  \bibinfo{person}{Aaron Maschinot}, \bibinfo{person}{Ce Liu}, {and}
  \bibinfo{person}{Dilip Krishnan}.} \bibinfo{year}{2020}\natexlab{}.
\newblock \showarticletitle{Supervised Contrastive Learning}. In
  \bibinfo{booktitle}{\emph{NeurIPS}}.
\newblock


\bibitem[Kipf and Welling(2017)]%
        {DBLP:conf/iclr/KipfW17}
\bibfield{author}{\bibinfo{person}{Thomas~N. Kipf} {and} \bibinfo{person}{Max
  Welling}.} \bibinfo{year}{2017}\natexlab{}.
\newblock \showarticletitle{Semi-Supervised Classification with Graph
  Convolutional Networks}. In \bibinfo{booktitle}{\emph{{ICLR} (Poster)}}.
  \bibinfo{publisher}{OpenReview.net}.
\newblock


\bibitem[Koren et~al\mbox{.}(2009)]%
        {DBLP:journals/computer/KorenBV09}
\bibfield{author}{\bibinfo{person}{Yehuda Koren}, \bibinfo{person}{Robert~M.
  Bell}, {and} \bibinfo{person}{Chris Volinsky}.}
  \bibinfo{year}{2009}\natexlab{}.
\newblock \showarticletitle{Matrix Factorization Techniques for Recommender
  Systems}.
\newblock \bibinfo{journal}{\emph{Computer}} \bibinfo{volume}{42},
  \bibinfo{number}{8} (\bibinfo{year}{2009}), \bibinfo{pages}{30--37}.
\newblock


\bibitem[Liang et~al\mbox{.}(2016)]%
        {DBLP:conf/www/LiangCMB16}
\bibfield{author}{\bibinfo{person}{Dawen Liang}, \bibinfo{person}{Laurent
  Charlin}, \bibinfo{person}{James McInerney}, {and} \bibinfo{person}{David~M.
  Blei}.} \bibinfo{year}{2016}\natexlab{}.
\newblock \showarticletitle{Modeling User Exposure in Recommendation}. In
  \bibinfo{booktitle}{\emph{{WWW}}}. \bibinfo{publisher}{{ACM}},
  \bibinfo{pages}{951--961}.
\newblock


\bibitem[Liang et~al\mbox{.}(2018)]%
        {DBLP:conf/www/LiangKHJ18}
\bibfield{author}{\bibinfo{person}{Dawen Liang}, \bibinfo{person}{Rahul~G.
  Krishnan}, \bibinfo{person}{Matthew~D. Hoffman}, {and} \bibinfo{person}{Tony
  Jebara}.} \bibinfo{year}{2018}\natexlab{}.
\newblock \showarticletitle{Variational Autoencoders for Collaborative
  Filtering}. In \bibinfo{booktitle}{\emph{{WWW}}}. \bibinfo{publisher}{{ACM}},
  \bibinfo{pages}{689--698}.
\newblock


\bibitem[Lin et~al\mbox{.}(2022)]%
        {DBLP:conf/www/LinTHZ22}
\bibfield{author}{\bibinfo{person}{Zihan Lin}, \bibinfo{person}{Changxin Tian},
  \bibinfo{person}{Yupeng Hou}, {and} \bibinfo{person}{Wayne~Xin Zhao}.}
  \bibinfo{year}{2022}\natexlab{}.
\newblock \showarticletitle{Improving Graph Collaborative Filtering with
  Neighborhood-enriched Contrastive Learning}. In
  \bibinfo{booktitle}{\emph{{WWW}}}. \bibinfo{publisher}{{ACM}},
  \bibinfo{pages}{2320--2329}.
\newblock


\bibitem[Liu et~al\mbox{.}(2021)]%
        {DBLP:conf/www/LiuCZGN21}
\bibfield{author}{\bibinfo{person}{Fan Liu}, \bibinfo{person}{Zhiyong Cheng},
  \bibinfo{person}{Lei Zhu}, \bibinfo{person}{Zan Gao}, {and}
  \bibinfo{person}{Liqiang Nie}.} \bibinfo{year}{2021}\natexlab{}.
\newblock \showarticletitle{Interest-aware Message-Passing {GCN} for
  Recommendation}. In \bibinfo{booktitle}{\emph{{WWW}}}.
  \bibinfo{publisher}{{ACM} / {IW3C2}}, \bibinfo{pages}{1296--1305}.
\newblock


\bibitem[Liu et~al\mbox{.}(2023)]%
        {DBLP:journals/corr/abs-2302-02113}
\bibfield{author}{\bibinfo{person}{Jiahao Liu}, \bibinfo{person}{Dongsheng Li},
  \bibinfo{person}{Hansu Gu}, \bibinfo{person}{Tun Lu}, \bibinfo{person}{Peng
  Zhang}, \bibinfo{person}{Li Shang}, {and} \bibinfo{person}{Ning Gu}.}
  \bibinfo{year}{2023}\natexlab{}.
\newblock \showarticletitle{Personalized Graph Signal Processing for
  Collaborative Filtering}.
\newblock \bibinfo{journal}{\emph{CoRR}}  \bibinfo{volume}{abs/2302.02113}
  (\bibinfo{year}{2023}).
\newblock


\bibitem[Liu et~al\mbox{.}(2022)]%
        {DBLP:conf/sigir/0001OM22}
\bibfield{author}{\bibinfo{person}{Siwei Liu}, \bibinfo{person}{Iadh Ounis},
  {and} \bibinfo{person}{Craig Macdonald}.} \bibinfo{year}{2022}\natexlab{}.
\newblock \showarticletitle{An MLP-based Algorithm for Efficient Contrastive
  Graph Recommendations}. In \bibinfo{booktitle}{\emph{{SIGIR}}}.
  \bibinfo{publisher}{{ACM}}, \bibinfo{pages}{2431--2436}.
\newblock


\bibitem[Ma et~al\mbox{.}(2019a)]%
        {DBLP:conf/icml/Ma0KW019}
\bibfield{author}{\bibinfo{person}{Jianxin Ma}, \bibinfo{person}{Peng Cui},
  \bibinfo{person}{Kun Kuang}, \bibinfo{person}{Xin Wang}, {and}
  \bibinfo{person}{Wenwu Zhu}.} \bibinfo{year}{2019}\natexlab{a}.
\newblock \showarticletitle{Disentangled Graph Convolutional Networks}. In
  \bibinfo{booktitle}{\emph{{ICML}}} \emph{(\bibinfo{series}{Proceedings of
  Machine Learning Research}, Vol.~\bibinfo{volume}{97})}.
  \bibinfo{publisher}{{PMLR}}, \bibinfo{pages}{4212--4221}.
\newblock


\bibitem[Ma et~al\mbox{.}(2019b)]%
        {DBLP:conf/nips/MaZ0Y019}
\bibfield{author}{\bibinfo{person}{Jianxin Ma}, \bibinfo{person}{Chang Zhou},
  \bibinfo{person}{Peng Cui}, \bibinfo{person}{Hongxia Yang}, {and}
  \bibinfo{person}{Wenwu Zhu}.} \bibinfo{year}{2019}\natexlab{b}.
\newblock \showarticletitle{Learning Disentangled Representations for
  Recommendation}. In \bibinfo{booktitle}{\emph{NeurIPS}}.
  \bibinfo{pages}{5712--5723}.
\newblock


\bibitem[Malitesta et~al\mbox{.}(2023)]%
        {DBLP:conf/um/MalitestaPANF23}
\bibfield{author}{\bibinfo{person}{Daniele Malitesta}, \bibinfo{person}{Claudio
  Pomo}, \bibinfo{person}{Vito~Walter Anelli}, \bibinfo{person}{Tommaso~Di
  Noia}, {and} \bibinfo{person}{Antonio Ferrara}.}
  \bibinfo{year}{2023}\natexlab{}.
\newblock \showarticletitle{An Out-of-the-Box Application for Reproducible
  Graph Collaborative Filtering extending the Elliot Framework}. In
  \bibinfo{booktitle}{\emph{{UMAP} (Adjunct Publication)}}.
  \bibinfo{publisher}{{ACM}}, \bibinfo{pages}{12--15}.
\newblock


\bibitem[Mao et~al\mbox{.}(2021a)]%
        {DBLP:conf/cikm/MaoZWDDXH21}
\bibfield{author}{\bibinfo{person}{Kelong Mao}, \bibinfo{person}{Jieming Zhu},
  \bibinfo{person}{Jinpeng Wang}, \bibinfo{person}{Quanyu Dai},
  \bibinfo{person}{Zhenhua Dong}, \bibinfo{person}{Xi Xiao}, {and}
  \bibinfo{person}{Xiuqiang He}.} \bibinfo{year}{2021}\natexlab{a}.
\newblock \showarticletitle{SimpleX: {A} Simple and Strong Baseline for
  Collaborative Filtering}. In \bibinfo{booktitle}{\emph{{CIKM}}}.
  \bibinfo{publisher}{{ACM}}, \bibinfo{pages}{1243--1252}.
\newblock


\bibitem[Mao et~al\mbox{.}(2021b)]%
        {DBLP:conf/cikm/MaoZXLWH21}
\bibfield{author}{\bibinfo{person}{Kelong Mao}, \bibinfo{person}{Jieming Zhu},
  \bibinfo{person}{Xi Xiao}, \bibinfo{person}{Biao Lu},
  \bibinfo{person}{Zhaowei Wang}, {and} \bibinfo{person}{Xiuqiang He}.}
  \bibinfo{year}{2021}\natexlab{b}.
\newblock \showarticletitle{UltraGCN: Ultra Simplification of Graph
  Convolutional Networks for Recommendation}. In
  \bibinfo{booktitle}{\emph{{CIKM}}}. \bibinfo{publisher}{{ACM}},
  \bibinfo{pages}{1253--1262}.
\newblock


\bibitem[Ouyang et~al\mbox{.}(2022)]%
        {DBLP:conf/cikm/OuyangWP22}
\bibfield{author}{\bibinfo{person}{Yi Ouyang}, \bibinfo{person}{Peng Wu}, {and}
  \bibinfo{person}{Li Pan}.} \bibinfo{year}{2022}\natexlab{}.
\newblock \showarticletitle{Asymmetrical Context-aware Modulation for
  Collaborative Filtering Recommendation}. In
  \bibinfo{booktitle}{\emph{{CIKM}}}. \bibinfo{publisher}{{ACM}},
  \bibinfo{pages}{1595--1604}.
\newblock


\bibitem[Paudel et~al\mbox{.}(2017)]%
        {DBLP:journals/tiis/PaudelCNB17}
\bibfield{author}{\bibinfo{person}{Bibek Paudel}, \bibinfo{person}{Fabian
  Christoffel}, \bibinfo{person}{Chris Newell}, {and} \bibinfo{person}{Abraham
  Bernstein}.} \bibinfo{year}{2017}\natexlab{}.
\newblock \showarticletitle{Updatable, Accurate, Diverse, and Scalable
  Recommendations for Interactive Applications}.
\newblock \bibinfo{journal}{\emph{{ACM} Trans. Interact. Intell. Syst.}}
  \bibinfo{volume}{7}, \bibinfo{number}{1} (\bibinfo{year}{2017}),
  \bibinfo{pages}{1:1--1:34}.
\newblock


\bibitem[Peng et~al\mbox{.}(2022a)]%
        {DBLP:conf/sigir/PengSM22}
\bibfield{author}{\bibinfo{person}{Shaowen Peng}, \bibinfo{person}{Kazunari
  Sugiyama}, {and} \bibinfo{person}{Tsunenori Mine}.}
  \bibinfo{year}{2022}\natexlab{a}.
\newblock \showarticletitle{Less is More: Reweighting Important Spectral Graph
  Features for Recommendation}. In \bibinfo{booktitle}{\emph{{SIGIR}}}.
  \bibinfo{publisher}{{ACM}}, \bibinfo{pages}{1273--1282}.
\newblock


\bibitem[Peng et~al\mbox{.}(2022b)]%
        {DBLP:conf/cikm/PengSM22}
\bibfield{author}{\bibinfo{person}{Shaowen Peng}, \bibinfo{person}{Kazunari
  Sugiyama}, {and} \bibinfo{person}{Tsunenori Mine}.}
  \bibinfo{year}{2022}\natexlab{b}.
\newblock \showarticletitle{{SVD-GCN:} {A} Simplified Graph Convolution
  Paradigm for Recommendation}. In \bibinfo{booktitle}{\emph{{CIKM}}}.
  \bibinfo{publisher}{{ACM}}, \bibinfo{pages}{1625--1634}.
\newblock


\bibitem[Perozzi et~al\mbox{.}(2014)]%
        {DBLP:conf/kdd/PerozziAS14}
\bibfield{author}{\bibinfo{person}{Bryan Perozzi}, \bibinfo{person}{Rami
  Al{-}Rfou}, {and} \bibinfo{person}{Steven Skiena}.}
  \bibinfo{year}{2014}\natexlab{}.
\newblock \showarticletitle{DeepWalk: online learning of social
  representations}. In \bibinfo{booktitle}{\emph{{KDD}}}.
  \bibinfo{publisher}{{ACM}}, \bibinfo{pages}{701--710}.
\newblock


\bibitem[Rao et~al\mbox{.}(2015)]%
        {DBLP:conf/nips/RaoYRD15}
\bibfield{author}{\bibinfo{person}{Nikhil Rao}, \bibinfo{person}{Hsiang{-}Fu
  Yu}, \bibinfo{person}{Pradeep Ravikumar}, {and} \bibinfo{person}{Inderjit~S.
  Dhillon}.} \bibinfo{year}{2015}\natexlab{}.
\newblock \showarticletitle{Collaborative Filtering with Graph Information:
  Consistency and Scalable Methods}. In \bibinfo{booktitle}{\emph{{NIPS}}}.
  \bibinfo{pages}{2107--2115}.
\newblock


\bibitem[Rao et~al\mbox{.}(2022)]%
        {DBLP:conf/kdd/RaoCLSYH22}
\bibfield{author}{\bibinfo{person}{Xuan Rao}, \bibinfo{person}{Lisi Chen},
  \bibinfo{person}{Yong Liu}, \bibinfo{person}{Shuo Shang},
  \bibinfo{person}{Bin Yao}, {and} \bibinfo{person}{Peng Han}.}
  \bibinfo{year}{2022}\natexlab{}.
\newblock \showarticletitle{Graph-Flashback Network for Next Location
  Recommendation}. In \bibinfo{booktitle}{\emph{{KDD}}}.
  \bibinfo{publisher}{{ACM}}, \bibinfo{pages}{1463--1471}.
\newblock


\bibitem[Rendle et~al\mbox{.}(2009)]%
        {DBLP:conf/uai/RendleFGS09}
\bibfield{author}{\bibinfo{person}{Steffen Rendle}, \bibinfo{person}{Christoph
  Freudenthaler}, \bibinfo{person}{Zeno Gantner}, {and} \bibinfo{person}{Lars
  Schmidt{-}Thieme}.} \bibinfo{year}{2009}\natexlab{}.
\newblock \showarticletitle{{BPR:} Bayesian Personalized Ranking from Implicit
  Feedback}. In \bibinfo{booktitle}{\emph{{UAI}}}. \bibinfo{publisher}{{AUAI}
  Press}, \bibinfo{pages}{452--461}.
\newblock


\bibitem[Resnick et~al\mbox{.}(1994)]%
        {DBLP:conf/cscw/ResnickISBR94}
\bibfield{author}{\bibinfo{person}{Paul Resnick}, \bibinfo{person}{Neophytos
  Iacovou}, \bibinfo{person}{Mitesh Suchak}, \bibinfo{person}{Peter Bergstrom},
  {and} \bibinfo{person}{John Riedl}.} \bibinfo{year}{1994}\natexlab{}.
\newblock \showarticletitle{GroupLens: An Open Architecture for Collaborative
  Filtering of Netnews}. In \bibinfo{booktitle}{\emph{{CSCW}}}.
  \bibinfo{publisher}{{ACM}}, \bibinfo{pages}{175--186}.
\newblock


\bibitem[Sarwar et~al\mbox{.}(2001)]%
        {DBLP:conf/www/SarwarKKR01}
\bibfield{author}{\bibinfo{person}{Badrul~Munir Sarwar},
  \bibinfo{person}{George Karypis}, \bibinfo{person}{Joseph~A. Konstan}, {and}
  \bibinfo{person}{John Riedl}.} \bibinfo{year}{2001}\natexlab{}.
\newblock \showarticletitle{Item-based collaborative filtering recommendation
  algorithms}. In \bibinfo{booktitle}{\emph{{WWW}}}.
  \bibinfo{publisher}{{ACM}}, \bibinfo{pages}{285--295}.
\newblock


\bibitem[Scarselli et~al\mbox{.}(2009)]%
        {DBLP:journals/tnn/ScarselliGTHM09}
\bibfield{author}{\bibinfo{person}{Franco Scarselli}, \bibinfo{person}{Marco
  Gori}, \bibinfo{person}{Ah~Chung Tsoi}, \bibinfo{person}{Markus
  Hagenbuchner}, {and} \bibinfo{person}{Gabriele Monfardini}.}
  \bibinfo{year}{2009}\natexlab{}.
\newblock \showarticletitle{The Graph Neural Network Model}.
\newblock \bibinfo{journal}{\emph{{IEEE} Trans. Neural Networks}}
  \bibinfo{volume}{20}, \bibinfo{number}{1} (\bibinfo{year}{2009}),
  \bibinfo{pages}{61--80}.
\newblock


\bibitem[Shen et~al\mbox{.}(2021)]%
        {DBLP:conf/cikm/ShenWZSZLL21}
\bibfield{author}{\bibinfo{person}{Yifei Shen}, \bibinfo{person}{Yongji Wu},
  \bibinfo{person}{Yao Zhang}, \bibinfo{person}{Caihua Shan},
  \bibinfo{person}{Jun Zhang}, \bibinfo{person}{Khaled~B. Letaief}, {and}
  \bibinfo{person}{Dongsheng Li}.} \bibinfo{year}{2021}\natexlab{}.
\newblock \showarticletitle{How Powerful is Graph Convolution for
  Recommendation?}. In \bibinfo{booktitle}{\emph{{CIKM}}}.
  \bibinfo{publisher}{{ACM}}, \bibinfo{pages}{1619--1629}.
\newblock


\bibitem[Steck(2013)]%
        {DBLP:conf/recsys/Steck13}
\bibfield{author}{\bibinfo{person}{Harald Steck}.}
  \bibinfo{year}{2013}\natexlab{}.
\newblock \showarticletitle{Evaluation of recommendations: rating-prediction
  and ranking}. In \bibinfo{booktitle}{\emph{RecSys}}.
  \bibinfo{publisher}{{ACM}}, \bibinfo{pages}{213--220}.
\newblock


\bibitem[Steck(2019)]%
        {DBLP:conf/www/Steck19}
\bibfield{author}{\bibinfo{person}{Harald Steck}.}
  \bibinfo{year}{2019}\natexlab{}.
\newblock \showarticletitle{Embarrassingly Shallow Autoencoders for Sparse
  Data}. In \bibinfo{booktitle}{\emph{{WWW}}}. \bibinfo{publisher}{{ACM}},
  \bibinfo{pages}{3251--3257}.
\newblock


\bibitem[Sun et~al\mbox{.}(2021)]%
        {DBLP:conf/www/SunCZPV21}
\bibfield{author}{\bibinfo{person}{Jianing Sun}, \bibinfo{person}{Zhaoyue
  Cheng}, \bibinfo{person}{Saba Zuberi}, \bibinfo{person}{Felipe P{\'{e}}rez},
  {and} \bibinfo{person}{Maksims Volkovs}.} \bibinfo{year}{2021}\natexlab{}.
\newblock \showarticletitle{{HGCF:} Hyperbolic Graph Convolution Networks for
  Collaborative Filtering}. In \bibinfo{booktitle}{\emph{{WWW}}}.
  \bibinfo{publisher}{{ACM} / {IW3C2}}, \bibinfo{pages}{593--601}.
\newblock


\bibitem[Sun et~al\mbox{.}(2020a)]%
        {DBLP:conf/kdd/SunGZZRHGTYHC20}
\bibfield{author}{\bibinfo{person}{Jianing Sun}, \bibinfo{person}{Wei Guo},
  \bibinfo{person}{Dengcheng Zhang}, \bibinfo{person}{Yingxue Zhang},
  \bibinfo{person}{Florence Regol}, \bibinfo{person}{Yaochen Hu},
  \bibinfo{person}{Huifeng Guo}, \bibinfo{person}{Ruiming Tang},
  \bibinfo{person}{Han Yuan}, \bibinfo{person}{Xiuqiang He}, {and}
  \bibinfo{person}{Mark Coates}.} \bibinfo{year}{2020}\natexlab{a}.
\newblock \showarticletitle{A Framework for Recommending Accurate and Diverse
  Items Using Bayesian Graph Convolutional Neural Networks}. In
  \bibinfo{booktitle}{\emph{{KDD}}}. \bibinfo{publisher}{{ACM}},
  \bibinfo{pages}{2030--2039}.
\newblock


\bibitem[Sun et~al\mbox{.}(2020c)]%
        {DBLP:conf/sigir/SunZGGTHMC20}
\bibfield{author}{\bibinfo{person}{Jianing Sun}, \bibinfo{person}{Yingxue
  Zhang}, \bibinfo{person}{Wei Guo}, \bibinfo{person}{Huifeng Guo},
  \bibinfo{person}{Ruiming Tang}, \bibinfo{person}{Xiuqiang He},
  \bibinfo{person}{Chen Ma}, {and} \bibinfo{person}{Mark Coates}.}
  \bibinfo{year}{2020}\natexlab{c}.
\newblock \showarticletitle{Neighbor Interaction Aware Graph Convolution
  Networks for Recommendation}. In \bibinfo{booktitle}{\emph{{SIGIR}}}.
  \bibinfo{publisher}{{ACM}}, \bibinfo{pages}{1289--1298}.
\newblock


\bibitem[Sun et~al\mbox{.}(2020b)]%
        {DBLP:conf/recsys/SunY00Q0G20}
\bibfield{author}{\bibinfo{person}{Zhu Sun}, \bibinfo{person}{Di Yu},
  \bibinfo{person}{Hui Fang}, \bibinfo{person}{Jie Yang},
  \bibinfo{person}{Xinghua Qu}, \bibinfo{person}{Jie Zhang}, {and}
  \bibinfo{person}{Cong Geng}.} \bibinfo{year}{2020}\natexlab{b}.
\newblock \showarticletitle{Are We Evaluating Rigorously? Benchmarking
  Recommendation for Reproducible Evaluation and Fair Comparison}. In
  \bibinfo{booktitle}{\emph{RecSys}}. \bibinfo{publisher}{{ACM}},
  \bibinfo{pages}{23--32}.
\newblock


\bibitem[Tang et~al\mbox{.}(2015)]%
        {DBLP:conf/www/TangQWZYM15}
\bibfield{author}{\bibinfo{person}{Jian Tang}, \bibinfo{person}{Meng Qu},
  \bibinfo{person}{Mingzhe Wang}, \bibinfo{person}{Ming Zhang},
  \bibinfo{person}{Jun Yan}, {and} \bibinfo{person}{Qiaozhu Mei}.}
  \bibinfo{year}{2015}\natexlab{}.
\newblock \showarticletitle{{LINE:} Large-scale Information Network Embedding}.
  In \bibinfo{booktitle}{\emph{{WWW}}}. \bibinfo{publisher}{{ACM}},
  \bibinfo{pages}{1067--1077}.
\newblock


\bibitem[Tao et~al\mbox{.}(2020)]%
        {DBLP:journals/ipm/TaoWWHHC20}
\bibfield{author}{\bibinfo{person}{Zhulin Tao}, \bibinfo{person}{Yinwei Wei},
  \bibinfo{person}{Xiang Wang}, \bibinfo{person}{Xiangnan He},
  \bibinfo{person}{Xianglin Huang}, {and} \bibinfo{person}{Tat{-}Seng Chua}.}
  \bibinfo{year}{2020}\natexlab{}.
\newblock \showarticletitle{{MGAT:} Multimodal Graph Attention Network for
  Recommendation}.
\newblock \bibinfo{journal}{\emph{Inf. Process. Manag.}} \bibinfo{volume}{57},
  \bibinfo{number}{5} (\bibinfo{year}{2020}), \bibinfo{pages}{102277}.
\newblock


\bibitem[van~den Berg et~al\mbox{.}(2017)]%
        {DBLP:journals/corr/BergKW17}
\bibfield{author}{\bibinfo{person}{Rianne van~den Berg},
  \bibinfo{person}{Thomas~N. Kipf}, {and} \bibinfo{person}{Max Welling}.}
  \bibinfo{year}{2017}\natexlab{}.
\newblock \showarticletitle{Graph Convolutional Matrix Completion}.
\newblock \bibinfo{journal}{\emph{CoRR}}  \bibinfo{volume}{abs/1706.02263}
  (\bibinfo{year}{2017}).
\newblock


\bibitem[Velickovic et~al\mbox{.}(2018)]%
        {DBLP:conf/iclr/VelickovicCCRLB18}
\bibfield{author}{\bibinfo{person}{Petar Velickovic}, \bibinfo{person}{Guillem
  Cucurull}, \bibinfo{person}{Arantxa Casanova}, \bibinfo{person}{Adriana
  Romero}, \bibinfo{person}{Pietro Li{\`{o}}}, {and} \bibinfo{person}{Yoshua
  Bengio}.} \bibinfo{year}{2018}\natexlab{}.
\newblock \showarticletitle{Graph Attention Networks}. In
  \bibinfo{booktitle}{\emph{{ICLR} (Poster)}}.
  \bibinfo{publisher}{OpenReview.net}.
\newblock


\bibitem[Wang et~al\mbox{.}(2019a)]%
        {DBLP:conf/kdd/Wang00LC19}
\bibfield{author}{\bibinfo{person}{Xiang Wang}, \bibinfo{person}{Xiangnan He},
  \bibinfo{person}{Yixin Cao}, \bibinfo{person}{Meng Liu}, {and}
  \bibinfo{person}{Tat{-}Seng Chua}.} \bibinfo{year}{2019}\natexlab{a}.
\newblock \showarticletitle{{KGAT:} Knowledge Graph Attention Network for
  Recommendation}. In \bibinfo{booktitle}{\emph{{KDD}}}.
  \bibinfo{publisher}{{ACM}}, \bibinfo{pages}{950--958}.
\newblock


\bibitem[Wang et~al\mbox{.}(2019b)]%
        {DBLP:conf/sigir/Wang0WFC19}
\bibfield{author}{\bibinfo{person}{Xiang Wang}, \bibinfo{person}{Xiangnan He},
  \bibinfo{person}{Meng Wang}, \bibinfo{person}{Fuli Feng}, {and}
  \bibinfo{person}{Tat{-}Seng Chua}.} \bibinfo{year}{2019}\natexlab{b}.
\newblock \showarticletitle{Neural Graph Collaborative Filtering}. In
  \bibinfo{booktitle}{\emph{{SIGIR}}}. \bibinfo{publisher}{{ACM}},
  \bibinfo{pages}{165--174}.
\newblock


\bibitem[Wang et~al\mbox{.}(2019c)]%
        {DBLP:journals/corr/abs-1905-08108}
\bibfield{author}{\bibinfo{person}{Xiang Wang}, \bibinfo{person}{Xiangnan He},
  \bibinfo{person}{Meng Wang}, \bibinfo{person}{Fuli Feng}, {and}
  \bibinfo{person}{Tat{-}Seng Chua}.} \bibinfo{year}{2019}\natexlab{c}.
\newblock \showarticletitle{Neural Graph Collaborative Filtering}.
\newblock \bibinfo{journal}{\emph{CoRR}}  \bibinfo{volume}{abs/1905.08108}
  (\bibinfo{year}{2019}).
\newblock


\bibitem[Wang et~al\mbox{.}(2020)]%
        {DBLP:conf/sigir/WangJZ0XC20}
\bibfield{author}{\bibinfo{person}{Xiang Wang}, \bibinfo{person}{Hongye Jin},
  \bibinfo{person}{An Zhang}, \bibinfo{person}{Xiangnan He},
  \bibinfo{person}{Tong Xu}, {and} \bibinfo{person}{Tat{-}Seng Chua}.}
  \bibinfo{year}{2020}\natexlab{}.
\newblock \showarticletitle{Disentangled Graph Collaborative Filtering}. In
  \bibinfo{booktitle}{\emph{{SIGIR}}}. \bibinfo{publisher}{{ACM}},
  \bibinfo{pages}{1001--1010}.
\newblock


\bibitem[Wang et~al\mbox{.}(2022a)]%
        {9736612}
\bibfield{author}{\bibinfo{person}{Xiangmeng Wang}, \bibinfo{person}{Qian Li},
  \bibinfo{person}{Dianer Yu}, \bibinfo{person}{Peng Cui},
  \bibinfo{person}{Zhichao Wang}, {and} \bibinfo{person}{Guandong Xu}.}
  \bibinfo{year}{2022}\natexlab{a}.
\newblock \showarticletitle{Causal Disentanglement for Semantics-Aware Intent
  Learning in Recommendation}.
\newblock \bibinfo{journal}{\emph{IEEE Transactions on Knowledge and Data
  Engineering}} (\bibinfo{year}{2022}), \bibinfo{pages}{1--1}.
\newblock
\urldef\tempurl%
\url{https://doi.org/10.1109/TKDE.2022.3159802}
\showDOI{\tempurl}


\bibitem[Wang et~al\mbox{.}(2022b)]%
        {DBLP:conf/wsdm/WangZS22}
\bibfield{author}{\bibinfo{person}{Zhenyi Wang}, \bibinfo{person}{Huan Zhao},
  {and} \bibinfo{person}{Chuan Shi}.} \bibinfo{year}{2022}\natexlab{b}.
\newblock \showarticletitle{Profiling the Design Space for Graph Neural
  Networks based Collaborative Filtering}. In
  \bibinfo{booktitle}{\emph{{WSDM}}}. \bibinfo{publisher}{{ACM}},
  \bibinfo{pages}{1109--1119}.
\newblock


\bibitem[Wei et~al\mbox{.}(2022b)]%
        {DBLP:conf/cikm/WeiLBL22}
\bibfield{author}{\bibinfo{person}{Chunyu Wei}, \bibinfo{person}{Jian Liang},
  \bibinfo{person}{Bing Bai}, {and} \bibinfo{person}{Di Liu}.}
  \bibinfo{year}{2022}\natexlab{b}.
\newblock \showarticletitle{Dynamic Hypergraph Learning for Collaborative
  Filtering}. In \bibinfo{booktitle}{\emph{{CIKM}}}.
  \bibinfo{publisher}{{ACM}}, \bibinfo{pages}{2108--2117}.
\newblock


\bibitem[Wei et~al\mbox{.}(2022a)]%
        {DBLP:conf/wsdm/WeiHXXZY22}
\bibfield{author}{\bibinfo{person}{Wei Wei}, \bibinfo{person}{Chao Huang},
  \bibinfo{person}{Lianghao Xia}, \bibinfo{person}{Yong Xu},
  \bibinfo{person}{Jiashu Zhao}, {and} \bibinfo{person}{Dawei Yin}.}
  \bibinfo{year}{2022}\natexlab{a}.
\newblock \showarticletitle{Contrastive Meta Learning with Behavior
  Multiplicity for Recommendation}. In \bibinfo{booktitle}{\emph{{WSDM}}}.
  \bibinfo{publisher}{{ACM}}, \bibinfo{pages}{1120--1128}.
\newblock


\bibitem[Wu et~al\mbox{.}(2021)]%
        {DBLP:conf/sigir/WuWF0CLX21}
\bibfield{author}{\bibinfo{person}{Jiancan Wu}, \bibinfo{person}{Xiang Wang},
  \bibinfo{person}{Fuli Feng}, \bibinfo{person}{Xiangnan He},
  \bibinfo{person}{Liang Chen}, \bibinfo{person}{Jianxun Lian}, {and}
  \bibinfo{person}{Xing Xie}.} \bibinfo{year}{2021}\natexlab{}.
\newblock \showarticletitle{Self-supervised Graph Learning for Recommendation}.
  In \bibinfo{booktitle}{\emph{{SIGIR}}}. \bibinfo{publisher}{{ACM}},
  \bibinfo{pages}{726--735}.
\newblock


\bibitem[Wu et~al\mbox{.}(2023)]%
        {DBLP:journals/csur/WuSZXC23}
\bibfield{author}{\bibinfo{person}{Shiwen Wu}, \bibinfo{person}{Fei Sun},
  \bibinfo{person}{Wentao Zhang}, \bibinfo{person}{Xu Xie}, {and}
  \bibinfo{person}{Bin Cui}.} \bibinfo{year}{2023}\natexlab{}.
\newblock \showarticletitle{Graph Neural Networks in Recommender Systems: {A}
  Survey}.
\newblock \bibinfo{journal}{\emph{{ACM} Comput. Surv.}} \bibinfo{volume}{55},
  \bibinfo{number}{5} (\bibinfo{year}{2023}), \bibinfo{pages}{97:1--97:37}.
\newblock


\bibitem[Xia et~al\mbox{.}(2022b)]%
        {DBLP:conf/www/XiaLGLLG22}
\bibfield{author}{\bibinfo{person}{Jiafeng Xia}, \bibinfo{person}{Dongsheng
  Li}, \bibinfo{person}{Hansu Gu}, \bibinfo{person}{Jiahao Liu},
  \bibinfo{person}{Tun Lu}, {and} \bibinfo{person}{Ning Gu}.}
  \bibinfo{year}{2022}\natexlab{b}.
\newblock \showarticletitle{{FIRE:} Fast Incremental Recommendation with Graph
  Signal Processing}. In \bibinfo{booktitle}{\emph{{WWW}}}.
  \bibinfo{publisher}{{ACM}}, \bibinfo{pages}{2360--2369}.
\newblock


\bibitem[Xia et~al\mbox{.}(2023)]%
        {DBLP:journals/corr/abs-2303-08537}
\bibfield{author}{\bibinfo{person}{Lianghao Xia}, \bibinfo{person}{Chao Huang},
  \bibinfo{person}{Jiao Shi}, {and} \bibinfo{person}{Yong Xu}.}
  \bibinfo{year}{2023}\natexlab{}.
\newblock \showarticletitle{Graph-less Collaborative Filtering}.
\newblock \bibinfo{journal}{\emph{CoRR}}  \bibinfo{volume}{abs/2303.08537}
  (\bibinfo{year}{2023}).
\newblock


\bibitem[Xia et~al\mbox{.}(2022a)]%
        {DBLP:conf/sigir/XiaHXZYH22}
\bibfield{author}{\bibinfo{person}{Lianghao Xia}, \bibinfo{person}{Chao Huang},
  \bibinfo{person}{Yong Xu}, \bibinfo{person}{Jiashu Zhao},
  \bibinfo{person}{Dawei Yin}, {and} \bibinfo{person}{Jimmy~X. Huang}.}
  \bibinfo{year}{2022}\natexlab{a}.
\newblock \showarticletitle{Hypergraph Contrastive Collaborative Filtering}. In
  \bibinfo{booktitle}{\emph{{SIGIR}}}. \bibinfo{publisher}{{ACM}},
  \bibinfo{pages}{70--79}.
\newblock


\bibitem[Yang et~al\mbox{.}(2018)]%
        {DBLP:conf/recsys/YangCWT18}
\bibfield{author}{\bibinfo{person}{Jheng{-}Hong Yang},
  \bibinfo{person}{Chih{-}Ming Chen}, \bibinfo{person}{Chuan{-}Ju Wang}, {and}
  \bibinfo{person}{Ming{-}Feng Tsai}.} \bibinfo{year}{2018}\natexlab{}.
\newblock \showarticletitle{HOP-rec: high-order proximity for implicit
  recommendation}. In \bibinfo{booktitle}{\emph{RecSys}}.
  \bibinfo{publisher}{{ACM}}, \bibinfo{pages}{140--144}.
\newblock


\bibitem[Yang et~al\mbox{.}(2022b)]%
        {DBLP:conf/www/000100LK22}
\bibfield{author}{\bibinfo{person}{Menglin Yang}, \bibinfo{person}{Min Zhou},
  \bibinfo{person}{Jiahong Liu}, \bibinfo{person}{Defu Lian}, {and}
  \bibinfo{person}{Irwin King}.} \bibinfo{year}{2022}\natexlab{b}.
\newblock \showarticletitle{{HRCF:} Enhancing Collaborative Filtering via
  Hyperbolic Geometric Regularization}. In \bibinfo{booktitle}{\emph{{WWW}}}.
  \bibinfo{publisher}{{ACM}}, \bibinfo{pages}{2462--2471}.
\newblock


\bibitem[Yang et~al\mbox{.}(2022a)]%
        {DBLP:conf/sigir/YangHXL22}
\bibfield{author}{\bibinfo{person}{Yuhao Yang}, \bibinfo{person}{Chao Huang},
  \bibinfo{person}{Lianghao Xia}, {and} \bibinfo{person}{Chenliang Li}.}
  \bibinfo{year}{2022}\natexlab{a}.
\newblock \showarticletitle{Knowledge Graph Contrastive Learning for
  Recommendation}. In \bibinfo{booktitle}{\emph{{SIGIR}}}.
  \bibinfo{publisher}{{ACM}}, \bibinfo{pages}{1434--1443}.
\newblock


\bibitem[Yao et~al\mbox{.}(2020)]%
        {DBLP:journals/corr/abs-2007-12865}
\bibfield{author}{\bibinfo{person}{Tiansheng Yao}, \bibinfo{person}{Xinyang
  Yi}, \bibinfo{person}{Derek~Zhiyuan Cheng}, \bibinfo{person}{Felix~X. Yu},
  \bibinfo{person}{Aditya~Krishna Menon}, \bibinfo{person}{Lichan Hong},
  \bibinfo{person}{Ed~H. Chi}, \bibinfo{person}{Steve Tjoa},
  \bibinfo{person}{Jieqi Kang}, {and} \bibinfo{person}{Evan Ettinger}.}
  \bibinfo{year}{2020}\natexlab{}.
\newblock \showarticletitle{Self-supervised Learning for Deep Models in
  Recommendations}.
\newblock \bibinfo{journal}{\emph{CoRR}}  \bibinfo{volume}{abs/2007.12865}
  (\bibinfo{year}{2020}).
\newblock


\bibitem[Ying et~al\mbox{.}(2018)]%
        {DBLP:conf/kdd/YingHCEHL18}
\bibfield{author}{\bibinfo{person}{Rex Ying}, \bibinfo{person}{Ruining He},
  \bibinfo{person}{Kaifeng Chen}, \bibinfo{person}{Pong Eksombatchai},
  \bibinfo{person}{William~L. Hamilton}, {and} \bibinfo{person}{Jure
  Leskovec}.} \bibinfo{year}{2018}\natexlab{}.
\newblock \showarticletitle{Graph Convolutional Neural Networks for Web-Scale
  Recommender Systems}. In \bibinfo{booktitle}{\emph{{KDD}}}.
  \bibinfo{publisher}{{ACM}}, \bibinfo{pages}{974--983}.
\newblock


\bibitem[Yu et~al\mbox{.}(2021)]%
        {DBLP:conf/www/YuYLWH021}
\bibfield{author}{\bibinfo{person}{Junliang Yu}, \bibinfo{person}{Hongzhi Yin},
  \bibinfo{person}{Jundong Li}, \bibinfo{person}{Qinyong Wang},
  \bibinfo{person}{Nguyen Quoc~Viet Hung}, {and} \bibinfo{person}{Xiangliang
  Zhang}.} \bibinfo{year}{2021}\natexlab{}.
\newblock \showarticletitle{Self-Supervised Multi-Channel Hypergraph
  Convolutional Network for Social Recommendation}. In
  \bibinfo{booktitle}{\emph{{WWW}}}. \bibinfo{publisher}{{ACM} / {IW3C2}},
  \bibinfo{pages}{413--424}.
\newblock


\bibitem[Yu et~al\mbox{.}(2022)]%
        {DBLP:conf/sigir/YuY00CN22}
\bibfield{author}{\bibinfo{person}{Junliang Yu}, \bibinfo{person}{Hongzhi Yin},
  \bibinfo{person}{Xin Xia}, \bibinfo{person}{Tong Chen},
  \bibinfo{person}{Lizhen Cui}, {and} \bibinfo{person}{Quoc Viet~Hung Nguyen}.}
  \bibinfo{year}{2022}\natexlab{}.
\newblock \showarticletitle{Are Graph Augmentations Necessary?: Simple Graph
  Contrastive Learning for Recommendation}. In
  \bibinfo{booktitle}{\emph{{SIGIR}}}. \bibinfo{publisher}{{ACM}},
  \bibinfo{pages}{1294--1303}.
\newblock


\bibitem[Yu and Qin(2020a)]%
        {DBLP:conf/icml/YuQ20}
\bibfield{author}{\bibinfo{person}{Wenhui Yu} {and} \bibinfo{person}{Zheng
  Qin}.} \bibinfo{year}{2020}\natexlab{a}.
\newblock \showarticletitle{Graph Convolutional Network for Recommendation with
  Low-pass Collaborative Filters}. In \bibinfo{booktitle}{\emph{{ICML}}}
  \emph{(\bibinfo{series}{Proceedings of Machine Learning Research},
  Vol.~\bibinfo{volume}{119})}. \bibinfo{publisher}{{PMLR}},
  \bibinfo{pages}{10936--10945}.
\newblock


\bibitem[Yu and Qin(2020b)]%
        {DBLP:conf/sigir/YuQ20}
\bibfield{author}{\bibinfo{person}{Wenhui Yu} {and} \bibinfo{person}{Zheng
  Qin}.} \bibinfo{year}{2020}\natexlab{b}.
\newblock \showarticletitle{Sampler Design for Implicit Feedback Data by
  Noisy-label Robust Learning}. In \bibinfo{booktitle}{\emph{{SIGIR}}}.
  \bibinfo{publisher}{{ACM}}, \bibinfo{pages}{861--870}.
\newblock


\bibitem[Zhang et~al\mbox{.}(2022)]%
        {DBLP:conf/sigir/ZhangL0WSLSZDZ22}
\bibfield{author}{\bibinfo{person}{Yiding Zhang}, \bibinfo{person}{Chaozhuo
  Li}, \bibinfo{person}{Xing Xie}, \bibinfo{person}{Xiao Wang},
  \bibinfo{person}{Chuan Shi}, \bibinfo{person}{Yuming Liu},
  \bibinfo{person}{Hao Sun}, \bibinfo{person}{Liangjie Zhang},
  \bibinfo{person}{Weiwei Deng}, {and} \bibinfo{person}{Qi Zhang}.}
  \bibinfo{year}{2022}\natexlab{}.
\newblock \showarticletitle{Geometric Disentangled Collaborative Filtering}. In
  \bibinfo{booktitle}{\emph{{SIGIR}}}. \bibinfo{publisher}{{ACM}},
  \bibinfo{pages}{80--90}.
\newblock


\bibitem[Zhao et~al\mbox{.}(2022)]%
        {DBLP:conf/sigir/0002WLCZDWSLW22}
\bibfield{author}{\bibinfo{person}{Minghao Zhao}, \bibinfo{person}{Le Wu},
  \bibinfo{person}{Yile Liang}, \bibinfo{person}{Lei Chen},
  \bibinfo{person}{Jian Zhang}, \bibinfo{person}{Qilin Deng},
  \bibinfo{person}{Kai Wang}, \bibinfo{person}{Xudong Shen},
  \bibinfo{person}{Tangjie Lv}, {and} \bibinfo{person}{Runze Wu}.}
  \bibinfo{year}{2022}\natexlab{}.
\newblock \showarticletitle{Investigating Accuracy-Novelty Performance for
  Graph-based Collaborative Filtering}. In \bibinfo{booktitle}{\emph{{SIGIR}}}.
  \bibinfo{publisher}{{ACM}}, \bibinfo{pages}{50--59}.
\newblock


\bibitem[Zheng et~al\mbox{.}(2018)]%
        {DBLP:conf/recsys/ZhengLJZY18}
\bibfield{author}{\bibinfo{person}{Lei Zheng}, \bibinfo{person}{Chun{-}Ta Lu},
  \bibinfo{person}{Fei Jiang}, \bibinfo{person}{Jiawei Zhang}, {and}
  \bibinfo{person}{Philip~S. Yu}.} \bibinfo{year}{2018}\natexlab{}.
\newblock \showarticletitle{Spectral collaborative filtering}. In
  \bibinfo{booktitle}{\emph{RecSys}}. \bibinfo{publisher}{{ACM}},
  \bibinfo{pages}{311--319}.
\newblock


\bibitem[Zhou et~al\mbox{.}(2022)]%
        {10.1145/3570501}
\bibfield{author}{\bibinfo{person}{Yuchen Zhou}, \bibinfo{person}{Yanan Cao},
  \bibinfo{person}{Yanmin Shang}, \bibinfo{person}{Chuan Zhou},
  \bibinfo{person}{Shirui Pan}, \bibinfo{person}{Zheng Lin}, {and}
  \bibinfo{person}{Qian Li}.} \bibinfo{year}{2022}\natexlab{}.
\newblock \showarticletitle{Explainable Hyperbolic Temporal Point Process for
  User-Item Interaction Sequence Generation}.
\newblock \bibinfo{journal}{\emph{ACM Trans. Inf. Syst.}} (\bibinfo{date}{nov}
  \bibinfo{year}{2022}).
\newblock
\showISSN{1046-8188}
\urldef\tempurl%
\url{https://doi.org/10.1145/3570501}
\showDOI{\tempurl}


\bibitem[Zhu et~al\mbox{.}(2022)]%
        {DBLP:conf/sigir/ZhuDSMLCXZ22}
\bibfield{author}{\bibinfo{person}{Jieming Zhu}, \bibinfo{person}{Quanyu Dai},
  \bibinfo{person}{Liangcai Su}, \bibinfo{person}{Rong Ma},
  \bibinfo{person}{Jinyang Liu}, \bibinfo{person}{Guohao Cai},
  \bibinfo{person}{Xi Xiao}, {and} \bibinfo{person}{Rui Zhang}.}
  \bibinfo{year}{2022}\natexlab{}.
\newblock \showarticletitle{{BARS:} Towards Open Benchmarking for Recommender
  Systems}. In \bibinfo{booktitle}{\emph{{SIGIR}}}. \bibinfo{publisher}{{ACM}},
  \bibinfo{pages}{2912--2923}.
\newblock


\bibitem[Ziegler et~al\mbox{.}(2005)]%
        {DBLP:conf/www/ZieglerMKL05}
\bibfield{author}{\bibinfo{person}{Cai{-}Nicolas Ziegler},
  \bibinfo{person}{Sean~M. McNee}, \bibinfo{person}{Joseph~A. Konstan}, {and}
  \bibinfo{person}{Georg Lausen}.} \bibinfo{year}{2005}\natexlab{}.
\newblock \showarticletitle{Improving recommendation lists through topic
  diversification}. In \bibinfo{booktitle}{\emph{{WWW}}}.
  \bibinfo{publisher}{{ACM}}, \bibinfo{pages}{22--32}.
\newblock


\end{thebibliography}


\end{document}